\documentclass[journal]{IEEEtran}

\usepackage{graphicx}
\usepackage{dcolumn}
\usepackage{bm}
\usepackage{epsfig}
\usepackage{float} 
\usepackage{amsmath}
\usepackage{amssymb}
\usepackage{tabularx}

\usepackage{xcolor}
\usepackage{hyperref}
\usepackage{cleveref}

\usepackage{cite}
\ifCLASSOPTIONcompsoc
 \usepackage[caption=false,font=normalsize,labelfont=sf,textfont=sf]{subfig}
\else
 \usepackage[caption=false,font=footnotesize]{subfig}
\fi
\usepackage{stfloats}
\usepackage{comment}
\usepackage{longtable}
\usepackage{url}

\usepackage{xspace}

\newcommand{\AuAu}{\mbox{Au$+$Au}\xspace}

\newcommand{\geant}{\mbox{\sc Geant4}\xspace}

\begin{document}


\title{Design and Beam Test Results for the sPHENIX Electromagnetic and Hadronic Calorimeter Prototypes}

\author{
  C.A.~Aidala, V.~Bailey,  S.~Beckman, R.~Belmont,  C.~Biggs,  J.~Blackburn, S.~Boose,  M.~Chiu,  M.~Connors, E.~Desmond, A.~Franz,  J.S.~Haggerty,
  X.~He,  M.M.~Higdon, J.~Huang,  K.~Kauder,  E.~Kistenev,  J.~LaBounty, J.G.~Lajoie,  M.~Lenz,  W.~Lenz,  S.~Li, V.R.~Loggins,  E.J.~Mannel, T.~Majoros,
  M.P.~McCumber,  J.L.~Nagle,  M.~Phipps,  C.~Pinkenburg,  S.~Polizzo,  C.~Pontieri, M.L.~Purschke, J.~Putschke, M.~Sarsour,  T.~Rinn, R.~Ruggiero,
  A.~Sen,  A.M.~Sickles,  M.J.~Skoby,  J.~Smiga, P.~Sobel, P.~Stankus, S.~Stoll,  A.~Sukhanov, E.~Thorsland, F.~Toldo,  R.S.~Towell, B.~Ujvari,  S.~Vazquez-Carson,  C.L.~Woody
  \thanks{Please see Acknowledgements for author affiliations.}
}

\maketitle

\begin{abstract}

The super Pioneering High Energy Nuclear Interaction eXperiment (sPHENIX) at the Relativistic Heavy Ion Collider (RHIC) will perform high precision measurements of jets and heavy flavor observables for a wide selection of nuclear collision systems, elucidating the microscopic nature of strongly interacting matter ranging from nucleons to the strongly coupled quark-gluon plasma. A prototype of the sPHENIX calorimeter system was
tested at the Fermilab Test Beam Facility as experiment T-1044 in the spring of 2016.  The
electromagnetic calorimeter (EMCal) prototype is composed of scintillating fibers embedded
in a mixture of tungsten powder and epoxy. The hadronic calorimeter (HCal) prototype is composed of tilted
steel plates alternating with plastic scintillator.  Results of the test beam reveal the
energy resolution for electrons in the EMCal is $2.8\%\oplus~15.5\%/\sqrt{E}$  and the
energy resolution for hadrons in the combined EMCal plus HCal system is $13.5\%\oplus
64.9\%/\sqrt{E}$. These results demonstrate that the performance of the proposed
calorimeter system satisfies the sPHENIX
specifications.

\end{abstract}

\begin{IEEEkeywords}
Calorimeters, electromagnetic calorimetry,
hadronic calorimetry, performance evaluation, prototypes, Relativistic
Heavy Ion Collider (RHIC), silicon photomultiplier
(SiPM), simulation, "Spaghetti" Calorimeter (SPACAL),
super Pioneering High Energy Nuclear Interaction eXperiment
(sPHENIX)
\end{IEEEkeywords}



\section{Introduction}
\label{sec:introduction}

The super Pioneering High Energy Nuclear Interaction eXperiment (sPHENIX)
is a planned experiment~\cite{Adare:2015kwa} at the Relativistic Heavy Ion
Collider (RHIC).  RHIC is a highly versatile machine that collides a diverse array of
nuclear beams from protons to heavy-ions, and supports a very broad physics program for
the study of both hot and cold quantum chromodynamics (QCD) matter.  sPHENIX is specifically designed for the
measurements of jets, quarkonia, and other rare processes originating from hard
scatterings to study the microscopic nature of strongly interacting matter ranging from
nucleons~\cite{Aschenauer:2016our} to the strongly coupled quark-gluon plasma
(QGP) created in collisions of gold ions at $\sqrt{s_{NN}}=$~200~GeV~\cite{Adcox:2004mh,Adams:2005dq,Back:2004je,Arsene:2004fa}. sPHENIX
is equipped with a tracking system and a three-segment calorimeter system, both of which have a full $2\pi$
acceptance in azimuth and a pseudorapidity coverage of $|\eta|<1.1$.  sPHENIX has acquired
the former BaBar magnet, which has an inner radius of 1.4~meters and an outer radius of
1.75~meters~\cite{OConnor:1998llb}.  The sPHENIX calorimeter system includes an
electromagnetic calorimeter and an inner hadronic calorimeter, which sit inside the
solenoid, and an outer hadronic calorimeter located outside of the magnet. The
electromagnetic calorimeter will be used for identifying photons, electrons and positrons. Photons
can be used to tag the energy of opposing jets traversing the QGP, while electrons and positrons will be
used to study quarkonia suppression and to tag heavy flavor jets. The combined EMCal and
HCal are used to measure the total electromagnetic and hadronic energy of jets, whose transverse energy range from 10--50~GeV. sPHENIX will be the first detector
at RHIC to employ hadronic calorimetry to enable full jet reconstruction at mid-rapidity.

The electromagnetic calorimeter (EMCal) design is based on both mechanical constraints and
physics requirements. The principal mechanical constraint for the EMCal is that it must be
compact, i.e. both the EMCal and the inner HCal must fit inside the solenoid magnet with enough space
remaining for a tracking system.  One major physics requirement is that it needs to have a
large solid angle with minimal inactive area to enable accurate jet measurements.  The
second major physics requirement is for the EMCal resolution and segmentation to be
compatible with the background conditions in heavy-ion collisions.  This means that a small
Moli\`{e}re radius and fine segmentation are required to reduce the influence of the
underlying heavy ion event background when measuring cluster energy of EM showers.

The most stringent requirement on the EMCal performance is that the energy resolution, when combined with track momentum information, should provide sufficient electron identification
to separate the upsilon signal from background.  The EMCal resolution
requirement for jets is less stringent.  In central \AuAu collisions with 0-10\%
centrality, the average EMCal energy from event background in a typical
EMCal tower cluster is 340 MeV~\cite{Adare:2015kwa}. Thus, an EMCal resolution of
$15\%/\sqrt{E}$ or better is sufficient to fulfill the sPHENIX physics requirements
of measuring photon and upsilon via their di-electron decay channels in
relativistic heavy ion collisions at $\sqrt{s_{NN}}=$~200~GeV.

The hadronic calorimeter (HCal) is a sampling calorimeter with two radial segments: one
inside the magnet and the other outside the magnet. The performance requirements of the
sPHENIX HCal are driven by the physics specifications related to measuring jets in
relativistic heavy ion collisions at $\sqrt{s_{NN}}=$~200~GeV.  At the jet energies of interest for the sPHENIX
physics program, the energy resolution in central \AuAu collisions 
is dominated by the underlying event, not the
energy resolution of the HCal~\cite{Adare:2015kwa}. 
The jet energy resolution needed for sPHENIX is $\sigma/E <120\%/\sqrt{E}$, 
which corresponds to an energy resolution for single hadrons in the full
calorimeter system to be $\sigma/E < 100\%/\sqrt{E}$.

Both the electromagnetic and hadronic calorimeters for sPHENIX are unique in their design in terms of other types of calorimeters that have been built in the past. The EMCal is a so-called SciFi (Scintillating fiber) "Spaghetti" Calorimeter (SPACAL), similar to those which have been used in other experiments ~\cite{Leverington:2008zz,Sedykh:2000ex,Armstrong:1998qs,Appuhn:1996na,Hertzog:1990md}. However, its design uses scintillating fibers embedded in a matrix of tungsten powder and epoxy, and will have a two dimensional tapered geometry that makes it approximately projective back to the interaction vertex in both $\eta$ and $\phi$. Both of these concepts required new and novel techniques in order to carry out its construction. The HCal has its absorber plates parallel to the beam direction, as opposed to being perpendicular to the direction of incident particles, as is typical for most other calorimeters ~\cite{CMS:JINST2008,ZEUS:Derrick1991,Bertolucci:1987zn}, and allows the outer steel plates to be used as a flux return for the solenoid magnet. The plates are also tilted, with opposite angles in the inner and outer HCals, in order to eliminate the possibility of particles passing through the calorimeter without encountering sufficient absorber (channeling). All of these features make the sPHENIX calorimeter system unique in terms of its overall design, and represent new developments in calorimetry for nuclear and high energy physics.

\begin{figure}[!hbt]
\begin{center}
\includegraphics[trim={12cm 21.5cm 10.2cm 4cm},clip, width=1.0\linewidth]{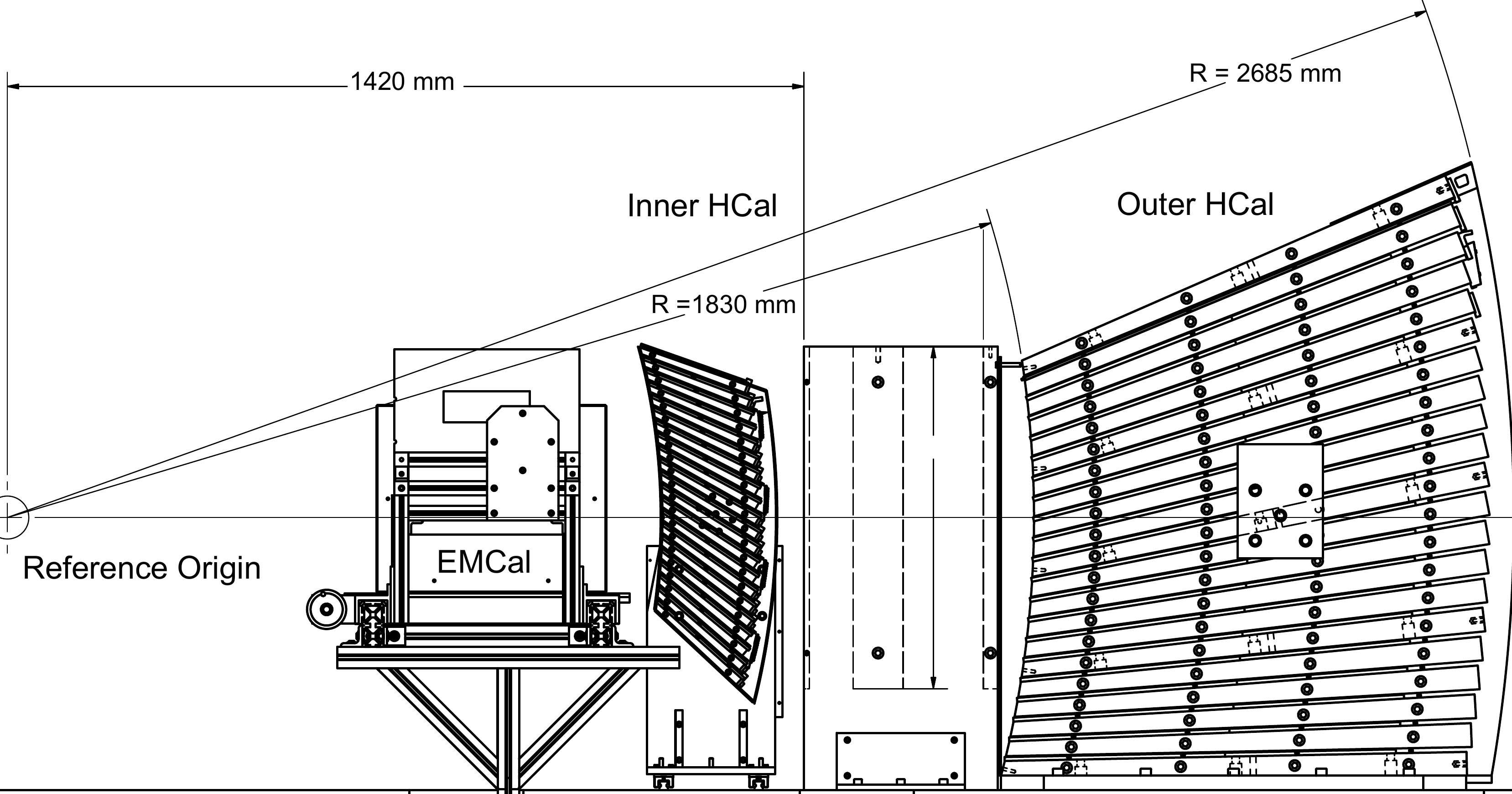}
\caption{Schematic diagram of the beam test setup. From left to right, it includes the EMCal,
inner HCal, mock cryostat, and outer HCal prototypes.
}
\label{fig:setup}
\end{center}
\end{figure}

To verify the design performance, a prototype of the sPHENIX calorimeter system was
assembled at Brookhaven National Laboratory and tested at the Fermilab Test Beam Facility
(FTBF) as experiment T-1044. A schematic diagram of the T-1044 test beam setup, including
the EMCal and HCal prototypes, is shown in Figure~\ref{fig:setup}. The beam goes from left
to right in the diagram interacting with the EMCal, the inner HCal, a ``mock cryostat" and
the outer HCal.  The mock cryostat, comprising three vertical plates of aluminum, is
placed between the inner and outer HCals to provide as many radiation lengths of
material as a particle would encounter traversing the sPHENIX solenoid (approximately 1.4~$X_0$). This article
presents the design of the EMCal and HCal prototypes as well as the results from the T-1044
experiment and simulations.

\section{Prototype Electromagnetic Calorimeter}
\label{sec:emcal}

\begin{figure}[!hbt]
\centering
\includegraphics[width=\linewidth]{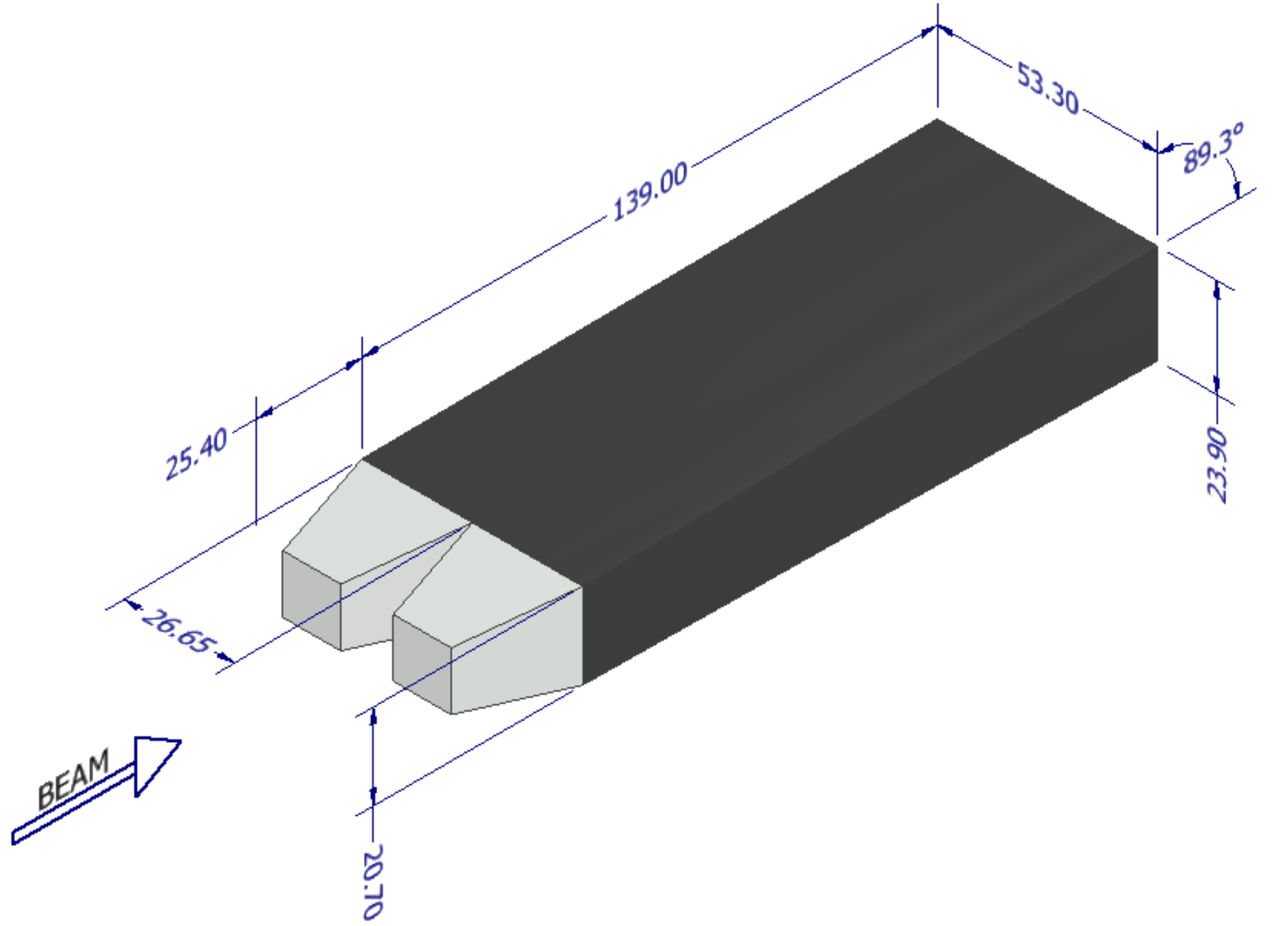}
\caption{\label{fig:fig_1DMod} 
A SPACAL block (black) with two light guides (gray) attached. Each light guide collects light from a half of the block, and forms a SPACAL tower. All dimension numbers have a unit of mm.
}
\end{figure}

The EMCal tower design consists of scintillating fibers embedded in the absorber material,
which is a matrix of tungsten powder infused with epoxy (W/SciFi).
This prototype is based on a design by a group at the University of California Los Angeles (UCLA)~\cite{Tsai:2012cpa,Tsai:2015bna} and
is similar to the SPACAL design used in a number of
experiments~\cite{Leverington:2008zz,Sedykh:2000ex,Armstrong:1998qs,Appuhn:1996na,Hertzog:1990md}.
The EMCal towers are designed in ``blocks'', with two towers composing one block.
The blocks are tapered in one-dimension, ($\phi$), as shown in Figure~\ref{fig:fig_1DMod} for this prototype, which is representative of the sPHENIX calorimeter at central rapidity. For larger rapidities, the blocks will be tapered in two dimensions ($\eta$ and $\phi$). The 2D tapered blocks will be studied in a subsequent beam test which is planned for early 2017.

The back of the two-tower block has a height of 2.39~cm,
the front has a height of 2.07~cm, and the total length of the two-tower blocks is 13.9~cm, which corresponds to about 18 radiation length.
1560 scintillating fibers extending along the length direction are embedded in the block. 
The outer diameter of the fibers is $0.47$~mm and the fibers are
arranged in a $30\times52$  triangular lattice with a nominal center-to-center spacing of approximately 1.0~mm.
At each of the four sides along the length of the block, the outer most fiber are kept approximately 0.5~mm away from the block surface, forming a skin of absorber to protect the outer fibers from being damaged during the manufacturing process. 
The density of the whole SPACAL block is approximately 10~g/cm$^3$, which is about half the density of metallic tungsten. The sampling fraction for EM-showers is about
2.3\% and the radiation length $X_0\approx$~0.7--0.8~cm.

The EMCal prototype consists of 32 two-tower blocks for a total of 64 towers.  Each tower
is equipped with a light guide on the front face and is read out by four silicon
photomultipliers (SiPMs) passively summed into a single preamp/electronics channel. Compared to the full sPHENIX EMCal covering $-1.1<\eta<1.1$ and full azimuth, this prototype represents a subset of towers covering $\Delta\eta\times\Delta\phi = 0.2\times0.2$ at mid-rapidity.

\subsection{EMCal Block Production}
\begin{figure}[!hbt]
\centering
\includegraphics[width=0.85\linewidth]{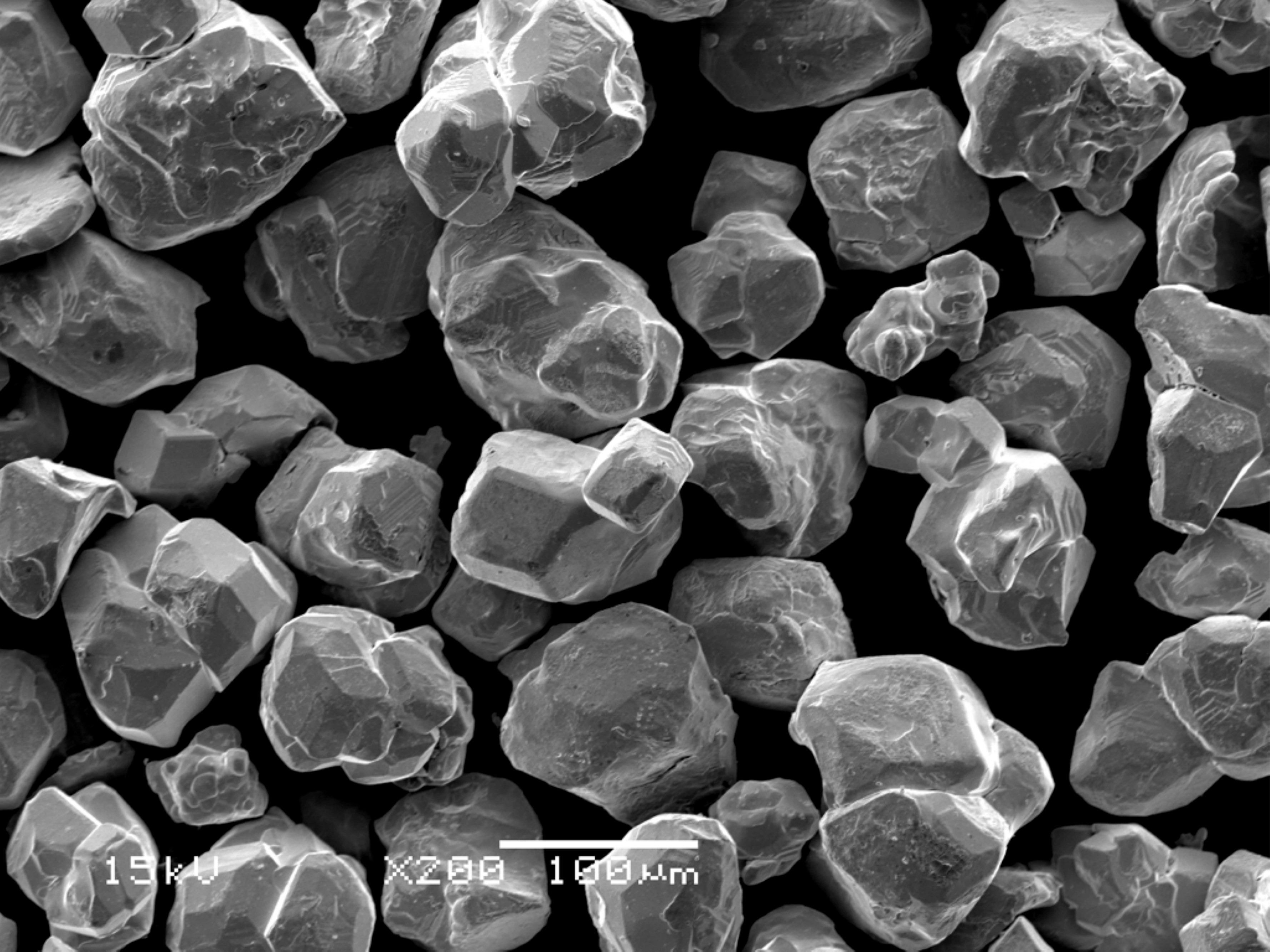}
\caption{\label{fig:fig_Tungsten}Tungsten powder was imaged using a JEOL 6060LV General Purpose Scanning Electron Microscope. The powder is sold as Technon Tungsten Powder -100 mesh, with 90\% of the particles ranging in size between 25-150 um. This wide distribution allows for higher packing density within each block.}
\end{figure}

The materials used in the EMCal block production are described in detail in
Table~\ref{tab:emcalmodule_materials}.  The tungsten powder used in these blocks comes
from Tungsten Heavy Powder Inc., San Diego (THP) and contains small amounts of alloy
material. Using a helium pycnometer, THP placed a lower
limit on the purity of the tungsten powder at 95.4\%.
An image of the typical powder particles is shown in Figure~\ref{fig:fig_Tungsten}.

The EMCal blocks were produced at two sites, THP and University of Illinois at
Urbana-Champaign (UIUC).  To produce the EMCal blocks, the scintillating fibers are placed
inside brass mesh screens, which position the fibers in a triangular pattern
with a nominal center-to-center spacing of approximately 1.0~mm. 
The screens are then separated longitudinally, placed in the mold, and
tilted to form the taper in one dimension. Tungsten powder is poured uniformly into the
mold and then epoxy is poured into the tungsten-fiber matrix. To aid the flow and
distribution of epoxy, a light vacuum is applied to the mold at the UIUC production site,
while THP used a centrifuge to distribute the epoxy.  After 24 hours, the SPACAL block is released from the
mold.  The blocks are first trimmed with carbide tipped cutters and then with diamond tipped
ones.  This allows the ends of the blocks to be cut without
degrading the light output of the fibers.

\begin{table}[!hbt]
\centering
\caption{EMCal block component materials}
\begin{tabular}{llr}
\hline
\multicolumn{1}{c}{\bf Material} & \multicolumn{1}{c}{\bf Property} & \multicolumn{1}{c}{\bf Value} \\
\hline
Tungsten powder & THP Technon 100 mesh	&  \\
& particle size	&  25-150 $\mu$m \\
& bulk density (solid) & $\ge$ 18.50 g/cm$^3$ \\
& tap density (powder) & $\ge$ 11.25 g/cm$^3$ \\
& purity & $\ge$ 95.4\% W \\
& impurities ($\le$ 5 percent) & Fe, Ni, O2, Co,  \\
&                              & Cr, Cu, Mo \\
Scintillating fiber 		& Kuraray SCSF78  & \\
                 		& (single cladding, blue) & \\
Epoxy                       & EPO-TEK 301 & \\
\hline
\end{tabular}
\label{tab:emcalmodule_materials}
\end{table}

\subsection{\label{sec:lightguide}Light Collection}

\begin{figure}[!hbt]
\begin{center}
\includegraphics[width=0.9\linewidth]{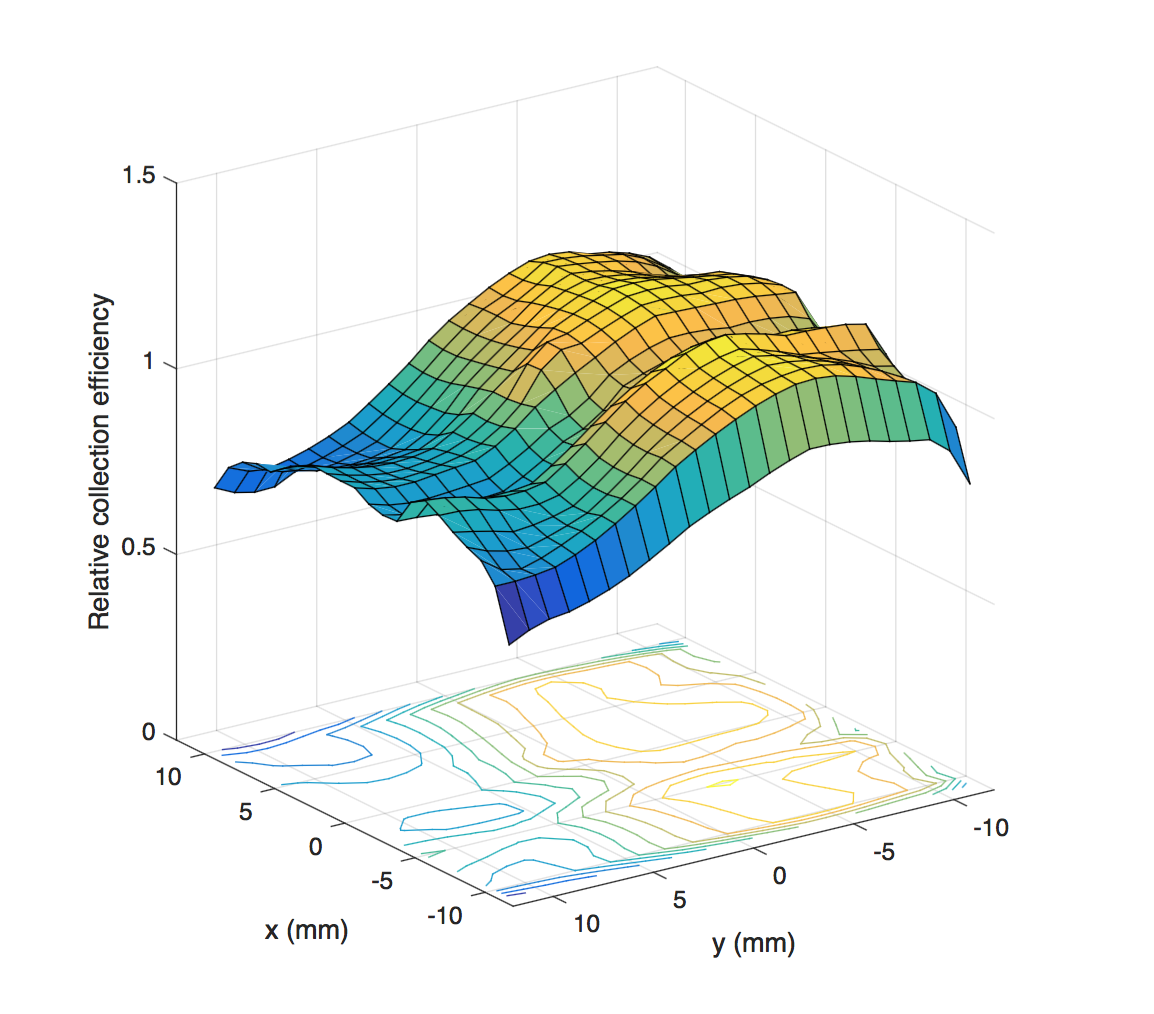} 
\end{center}
\caption{\label{fig:EMCal_LG_REL_EFFICIENCY} Relative light collection efficiency for the light guide and SiPM assembly with respect to the input fiber position in X-Y.}
\end{figure}

The EMCal light guide is a machined acrylic trapezoidal prism that fully covers
one tower of the W/SciFi two-tower block (2.64~$\times$~2.36~cm$^2$) and transitions over a 2.54~cm height to
an area (1.4~$\times$~1.4~cm$^2$) to accommodate a 2~$\times$~2 array of SiPMs.
The light guides are epoxied to the thin end of the two-tower block. Four SiPMs,
mounted on an EMCal preamp printed circuit board, are used to read out the block. The SiPMs are coupled to the
light guide using General Electric Silicones RTV615~\cite{RTV615TechnicalDatasheet}.

To measure the overall efficiency of the light guide, one tower of a W/SciFi block is
optically coupled to a 2 inch window photomultiplier tube (PMT). The PMT window fully
covers the readout surface of the block, and the readout end of the tower is masked.
The analog-to-digital converter (ADC) distribution arising from cosmic rays is measured
with trigger counters above and below
the block.  An acrylic light guide is then optically coupled between the block and the PMT,
and the measurement is repeated.  Relative to the directly coupled measurement, the light
guide measurement yields 71\% of the light, which represents the overall efficiency of the
light guide.

To map the uniformity of the light guide, a UV-pulse-excited scintillation fiber is scanned through the input end of the light guide and the response is read out using an array of 2~$\times$~2 SiPMs and preamplifier as in the prototype.
The measured relative collection efficiency with respect to the input fiber position in X-Y is shown in Figure~\ref{fig:EMCal_LG_REL_EFFICIENCY}. The center of the area bounded by the four SiPMs is offset from the center of the light guide, causing an asymmetry of the collection efficiency with respect to the center of the light guide.
Throughout the input cross section of the light guide, $\sim30\%$ relative variation is observed, which leads to $\sim20\%$ position dependent energy response variation for electromagnetic showers as discussed in Section \ref{sec:results}.


\subsection{Assembly}

\begin{figure}[!hbt]
\begin{center}
\includegraphics[width=1.0\linewidth]{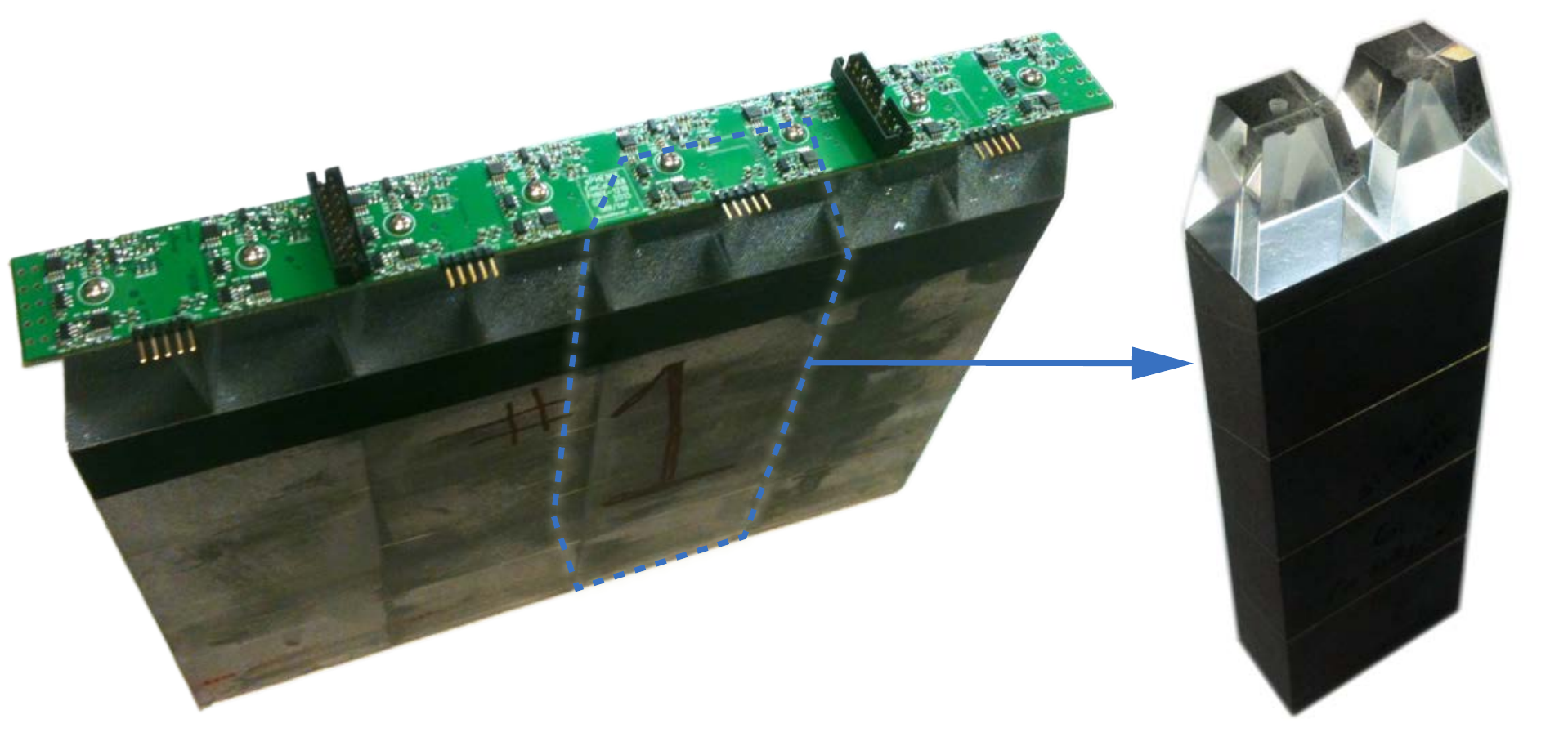}
\caption{A row of four EMCal blocks with light guides and preamp board on the left. Among them, one of the blocks and its two light guides are highlighted and displayed separately on the right side.}
\label{fig:EMCal_LG_preamp}
\end{center}
\end{figure}

After the blocks are produced at THP and UIUC, they are assembled at BNL prior to shipping the completed EMCal protoype to Fermilab for the test beam. The blocks are first epoxied together into
rows of eight towers in a gluing fixture, which aligns the front readout surface of the
blocks in a single plane. Two layers of Vikuiti Specular Reflector Film (ESR) reflective film~\cite{Vikuiti_ESR} are then epoxied to
the back surface of each of the rows. Light guides are epoxied to the front
surface of the row.
The preamplifier board, which carries four SiPMs per tower, is used to align the light guides on the towers.
The SiPMs are optically coupled to the light guides, and the
board carrying the SiPMs is mechanically secured by a screw to the center of each light guide, as shown in
Figure~\ref{fig:EMCal_LG_preamp}.
Eight rows of EMCal blocks are stacked and placed in a light-tight enclosure box.
The preamplifier heat output is 2.5 W/board, necessitating an active cooling system.
A blower is used to drive air through the enclosure box, providing sufficient cooling for the preamplifier and SiPM.

\section{Prototype Hadronic  Calorimeter}
\label{sec:hcal}

The inner and outer HCal prototypes are constructed as a small pseudorapidity and azimuthal segment ($\Delta\eta\times\Delta\phi = 0.4\times0.4$ at mid-rapidity) of the
full scale sPHENIX design, with alternating layers of scintillator tiles and steel
absorber plates.  The absorber plates are tapered and tilted from the radial direction to
provide more uniform sampling in azimuth.  Extruded tiles of plastic scintillator with an
embedded wavelength shifting (WLS) fiber are interspersed between the absorber plates. The tilt
angle is chosen so that a radial track from the center of the interaction region traverses
at least four scintillator tiles of each HCal.  Each tile is read out at the outer radius with SiPMs.
The analog signals from five tiles are summed to a single
preamplifier channel to form a single calorimeter tower.


\begin{figure}[!hbt]
  \subfloat[Scintillator tile production for inner HCal]{ \includegraphics[trim={0cm 0cm 0cm 0.8cm},clip,width=0.9\linewidth]{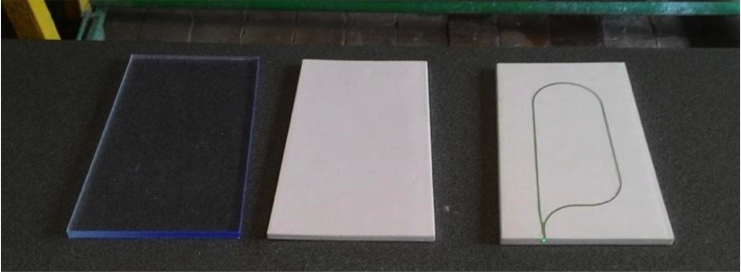}}
  \\
  \subfloat[Inner HCal tile design patterns]{ \includegraphics[trim={0cm 4.5cm 0cm 2.5cm},clip,width=1.0\linewidth]{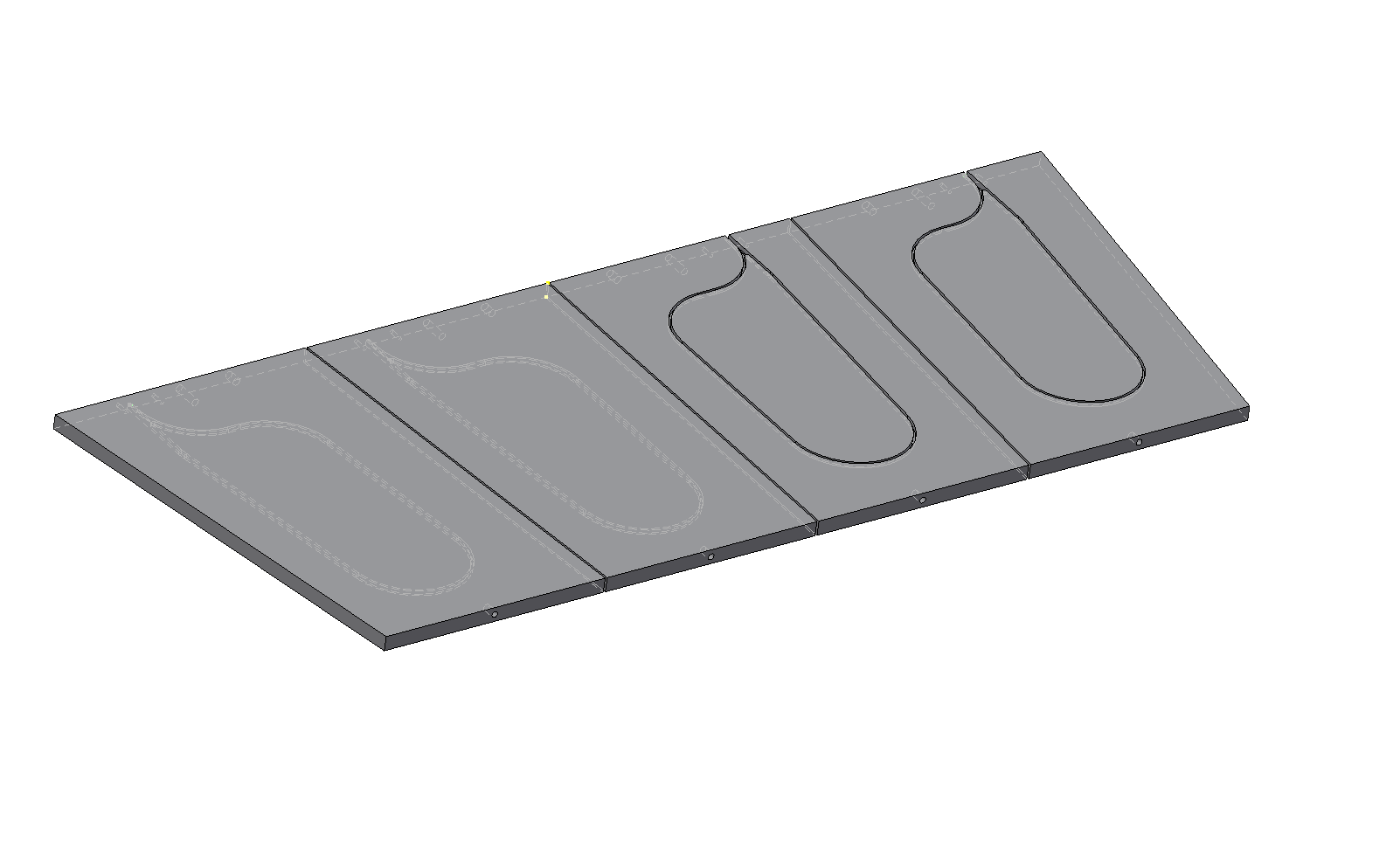} }
  \\
  \subfloat[Plastic coupler to attach the SiPM at the fiber exit]{ \includegraphics[trim={0cm 3cm 0cm 5cm},clip,width=1.0\linewidth]{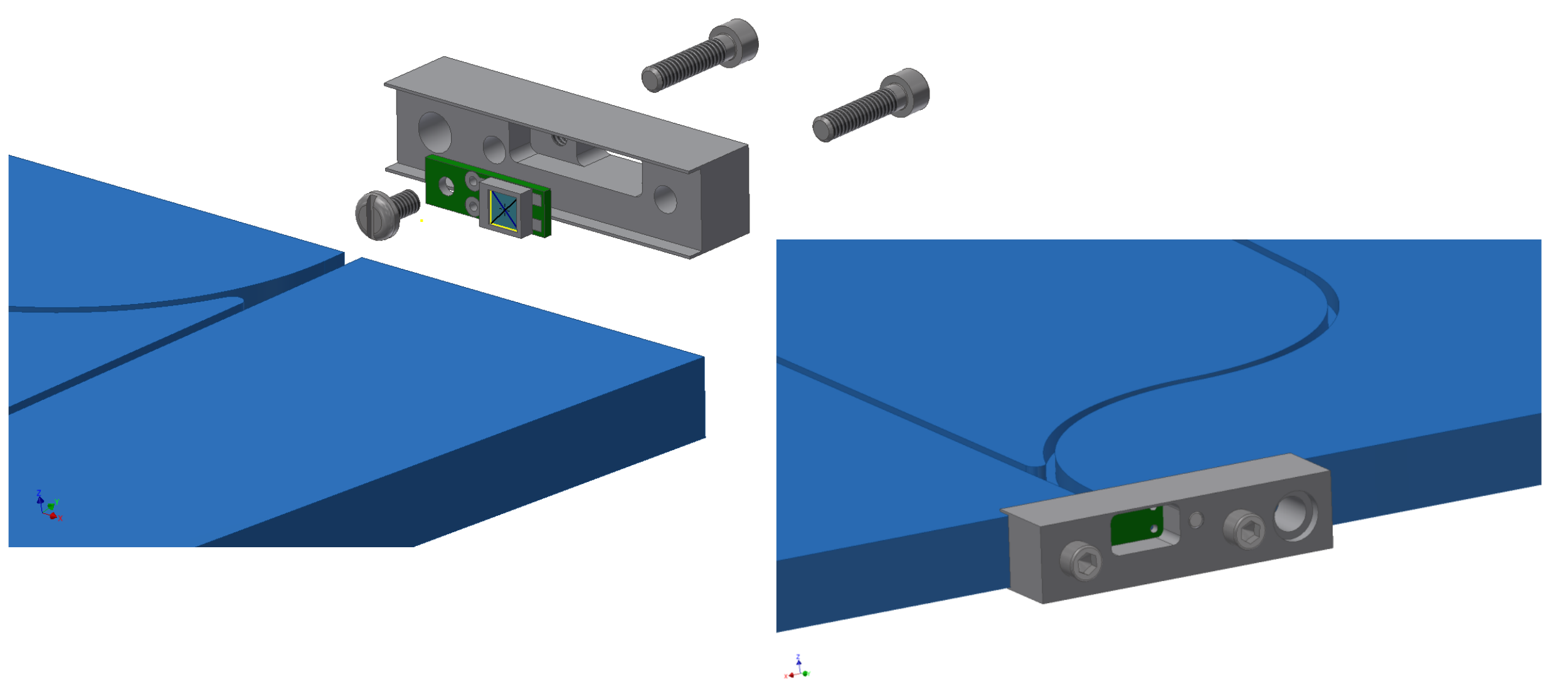} }
\caption{HCal tile production. (a) Inner HCal scintillating tiles in several stages of production. From left to right tiles are machined, then coated and embedded with WLS fiber.
(b) 4 scintillating tiles arranged symmetrically around $\eta$~=~0 to be inserted between the steel absorber plates.
(c) SiPM installation at the fiber exit using a plastic coupler.}
\label{fig:HCAL_tile_production}
\end{figure}

\begin{table}[!hbt]
\centering
\caption{Properties of the HCal scintillating tiles and fiber.}
\begin{tabular}{p{0.4\linewidth}p{0.4\linewidth}}
\hline
Property &  \\
\hline
Plastic                                & Extruded polystyrene \\
Scintillation dopant                   & 1.5\% of PTP and 0.01\% POPOP \\
Reflective coating                     & Proprietary coating by surface exposure to aromatic solvents\\
Reflective layer thickness             & 50~$\mu$m \\
Wrapping                               & 100~$\mu$m Al foil followed by one layer of 30~$\mu$m cling-wrap \\
                                       & and a 100~$\mu$m layer of black vinyl tape  \\
Attenuation length in lateral direction & Approximately~2-2.5~m \\
           (with respect to extrusion) &  \\
Wavelength shifting fiber              & Single clad Kuraray Y11 \\
Formulation                            & 200, K-27, S-Type\\
Cladding material                      & Polymethylmethacrylate (PMMA) \\
Fiber diameter                         & 1~mm \\
Emission peak                          & 476~nm\\
Fiber core attenuation length          & $>$~2~m \\
Optical cement                         & EPO-TEK 3015 \\
\hline
\end{tabular}
\label{tab:hcalscint}
\end{table}

\begin{figure}[!hbt]
 \begin{center}
 \includegraphics[trim={3cm 0cm 4cm 3.2cm},clip,width=1.\linewidth]{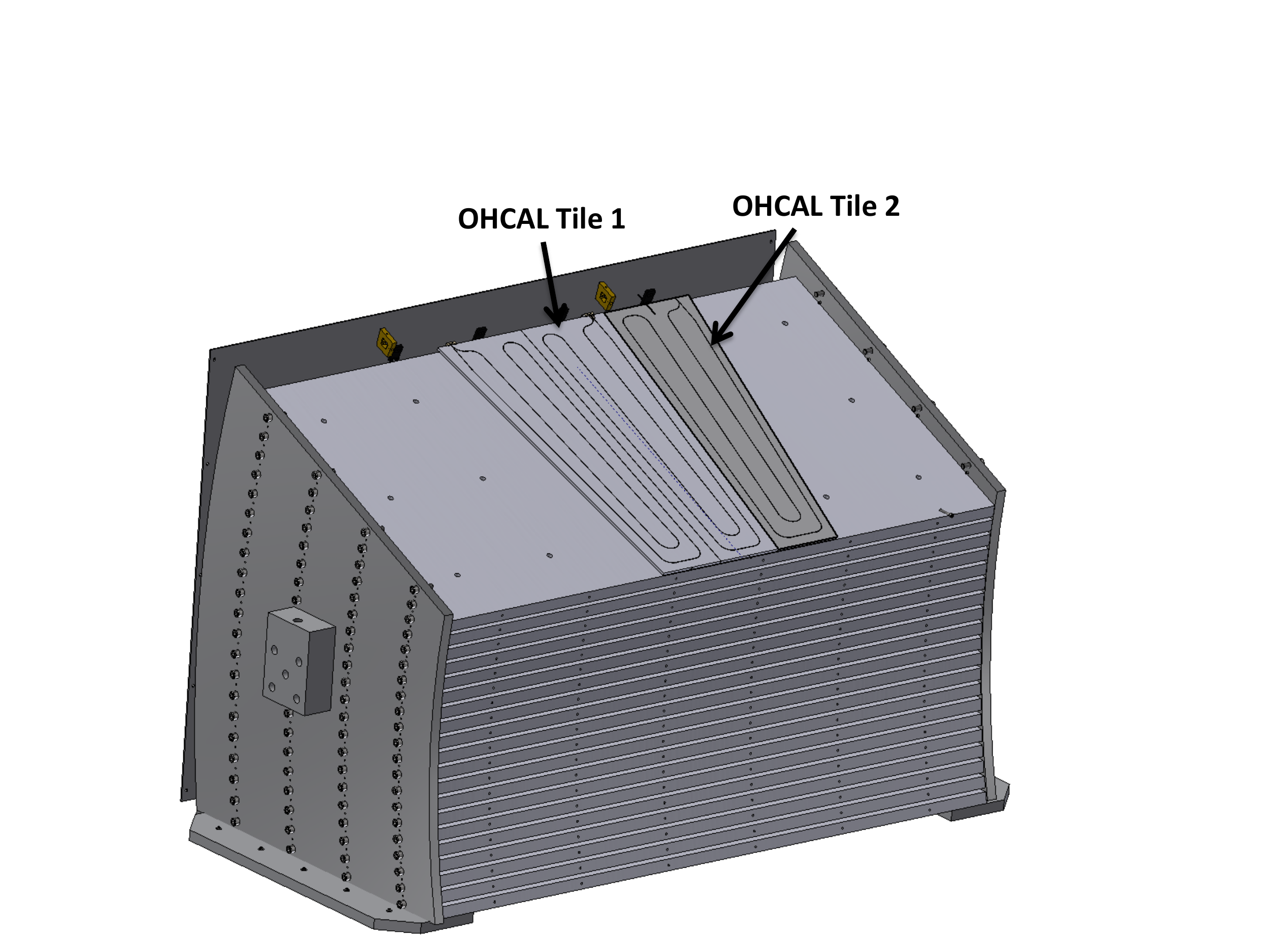}
 \end{center}
 \caption{Schematic diagram of the outer HCal tile designs and assembly. 20 steel absorber plates are stacked together, then 80 scintillating tiles are inserted between them. Tile fiber patterns are shown on the top tiles.}
 \label{fig:scint_tiles_outer}
 \end{figure}

The properties of the HCal scintillating tiles are listed in
Table~\ref{tab:hcalscint}. Figure~\ref{fig:HCAL_tile_production}~(a) shows the
steps of tile production.
Figure~\ref{fig:HCAL_tile_production}~(a,b) and
Figure~\ref{fig:scint_tiles_outer} show the inner and outer HCal fiber routing
patterns. The Kuraray~\cite{Kuraray} single-clad fiber is chosen due to its flexibility and longevity,
both of which are critical in the geometry with multiple fiber bends. The properties of
the HCal wavelength shifting fibers are included in Table~\ref{tab:hcalscint}. The
fiber routing is designed so that any energy deposited in the scintillator is within
a 2.5~cm distance from a WLS fiber, and the bend radius of any turn in the fiber has been limited to
2.5~cm to limit mechanical stress and light loss, based on the experience of the T2K
collaboration~\cite{Bryant:2007uwb} as well as the experience with the test tiles.

The scintillation light produced in the tiles by ionization from charged particles is kept
inside the tile and reflected diffusely by a reflective coating and reflective tile
wrapping.  The light is absorbed by the fiber embedded in the scintillator.  As shown in
Figure~\ref{fig:HCAL_tile_production} (c), the two ends of the fiber are brought to the
outer edge of the tile where a small plastic mount supports a 3~$\times$~3~mm$^2$
SiPM at the fiber exit.  The fiber exit is orthogonal to the tile edge and glued at a
depth in the tile that allows for installation of the SiPM centered around the fiber
exits.
The air gap between the fiber ends and the face of the SiPM allows the emitted light to
spread over the face of the SiPM, reducing the probability of optical saturation resulting
from the two or more photons impinging on the same pixel. A gap
of 0.75~mm satisfies the following two requirements: (1) there be no more than a 5\% variation in the SiPM response when
fibers and SiPM are misaligned by 0.2~mm; (2) no more than 20\% loss of the light outside
of SiPM sensitive area.

\subsection{Tile Construction}

Scintillating tiles for the calorimeter are manufactured by the UNIPLAST Company in
Vladimir, Russia. A dry mix of polystyrene granules, p-terphenyl (PTP), and 1,4-bis-2-(5-Phenyloxazolyl)-benzene (POPOP) is melted and
extruded, producing a continuous band of hot
scintillating plastic 25~cm wide.

The scintillator is then cut into 2~m long pieces which are inspected for defects and
discolorations and, if this low level control is passed, mechanically machined into the
tiles according to the specified dimensions. The tiles are then placed in a bath of aromatic
solvents resulting in the development of a white diffuse reflective coating over the whole
tile surface with an average thickness of 50~$\mu$m. This process also removes microscopic
non-uniformities normally present on the surface of extruded plastic, 
which decreases aging and improves the ability of the tile to withstand pressure without crazing. 
It also enhances the
efficiency of light collection in tiles with embedded fibers. Coated tiles are then
grooved and WLS fibers are embedded. The fibers are glued using optical epoxy (EPO-TEK 301) with special
care given to fiber positioning at the exit from the tile. The fibers are cut at the tile
edge and polished by hand.

\subsection{Tile Testing}

To determine the light response across the tiles, various studies have been performed. In
one setup, an LED with a collimator is attached to a mount on a two-dimensional rail
system with very accurate stepper motors.  This allows an automated analysis with very
high positional precision.  The LED scans of the outer HCal tiles consist of 174 points in
the long direction ($X$) and 54~points in the short direction ($Y$) for a total of 9,396~points.
The scan positions are 0.5~cm (approximately the LED spot size) apart in each direction.
The principal disadvantage of an LED scan is that light is inserted into the tile directly
rather than being induced by ionizing radiation.
During the FTBF test beam running, a ``tile mapper'' was constructed and placed on a
two-dimensional motion table.
The motion table
moves up/down and left/right, keeping the position along the beam direction fixed.  The
tile mapper included four outer HCal tiles placed perpendicular to the beam direction, so
that movement on the motion table corresponds to different positions on the tile face.
Each tile is read out individually, which enables a detailed study of the light response
as a function of position.  The scan consists of 20 total positions, 10 positions focused
on the inner part of the tile and 10 focused on the outer part of the tile.  A few of the
outer scan positions fall near the edge and are excluded from the analysis.  This study was 
performed with the 16 GeV negative pion beam.

\begin{figure*}[!hbt]
\begin{center}
\includegraphics[trim={1cm 0cm 1cm 0cm},clip,width=\linewidth]{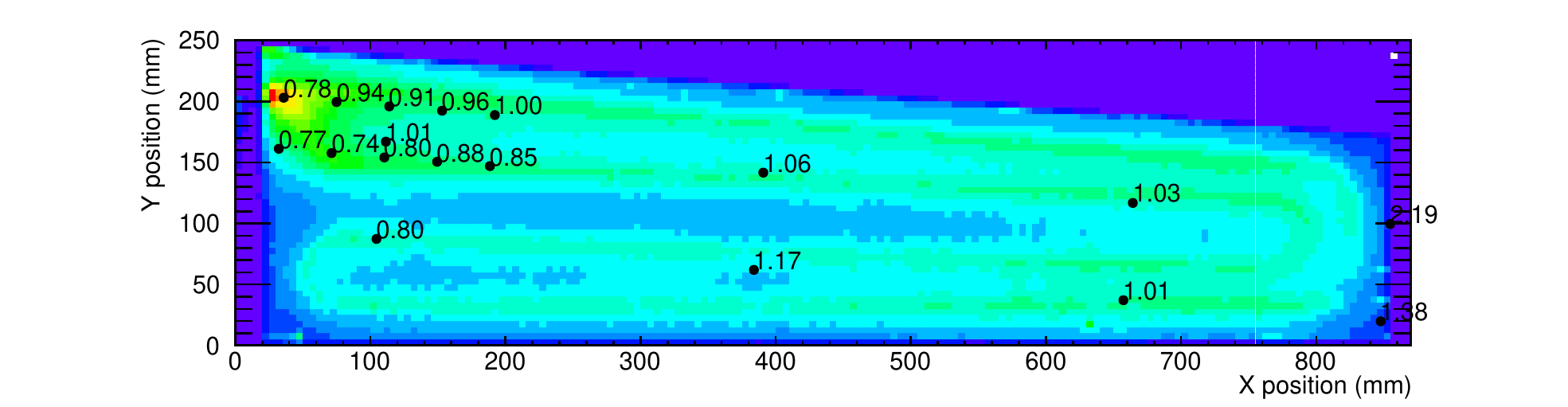}
\end{center}
\caption{LED response of a scintillation outer HCal tile with tile mapper scan data overlaid as black points.
The numerical value shown at each point is the normalized ratio of the LED response to the tile mapper response.
}
\label{fig:LED_map}
\end{figure*}

Figure~\ref{fig:LED_map} shows the LED scan of an outer HCal tile using 405 nm UV
LED. Additional scans were performed  using 375 and 361 nm UV LEDs with similar results. The overlaid black circles are the beam scan positions on a different tile.
The relative positional
accuracy of the points is 0.2-0.3 cm.  The numbers show the ratio of the average ADC value of
the 16 GeV pion data to the average ADC value of the LED scan for that point.  The
normalization is arbitrarily chosen so that the numerical values are near unity.

Most of the points have ratio values close to unity, indicating good agreement between
the 16 GeV pion data and the LED data.
The points close to the SiPM, which can be seen as the red region in the
upper left, show a downward trend in the ratio values, suggesting that the intense bright spot in the LED data
is not as intense in the 16 GeV pion data.  Additionally, the lower set of the
five inner points are systematically a little lower than the LED data, and they appear to be
right on top of the fiber.  This is most likely due to the fact that, in the LED scan,
some of the light from the LED is captured directly by the fiber, so there is a modest
enhancement directly over the fiber that is naturally not present in the 16 GeV pion data.
Both sets of five inner points, however, show a decreasing trend as the points get close to the
SiPM.

\begin{figure}[!hbt]
\begin{center}
\includegraphics[width=\linewidth]{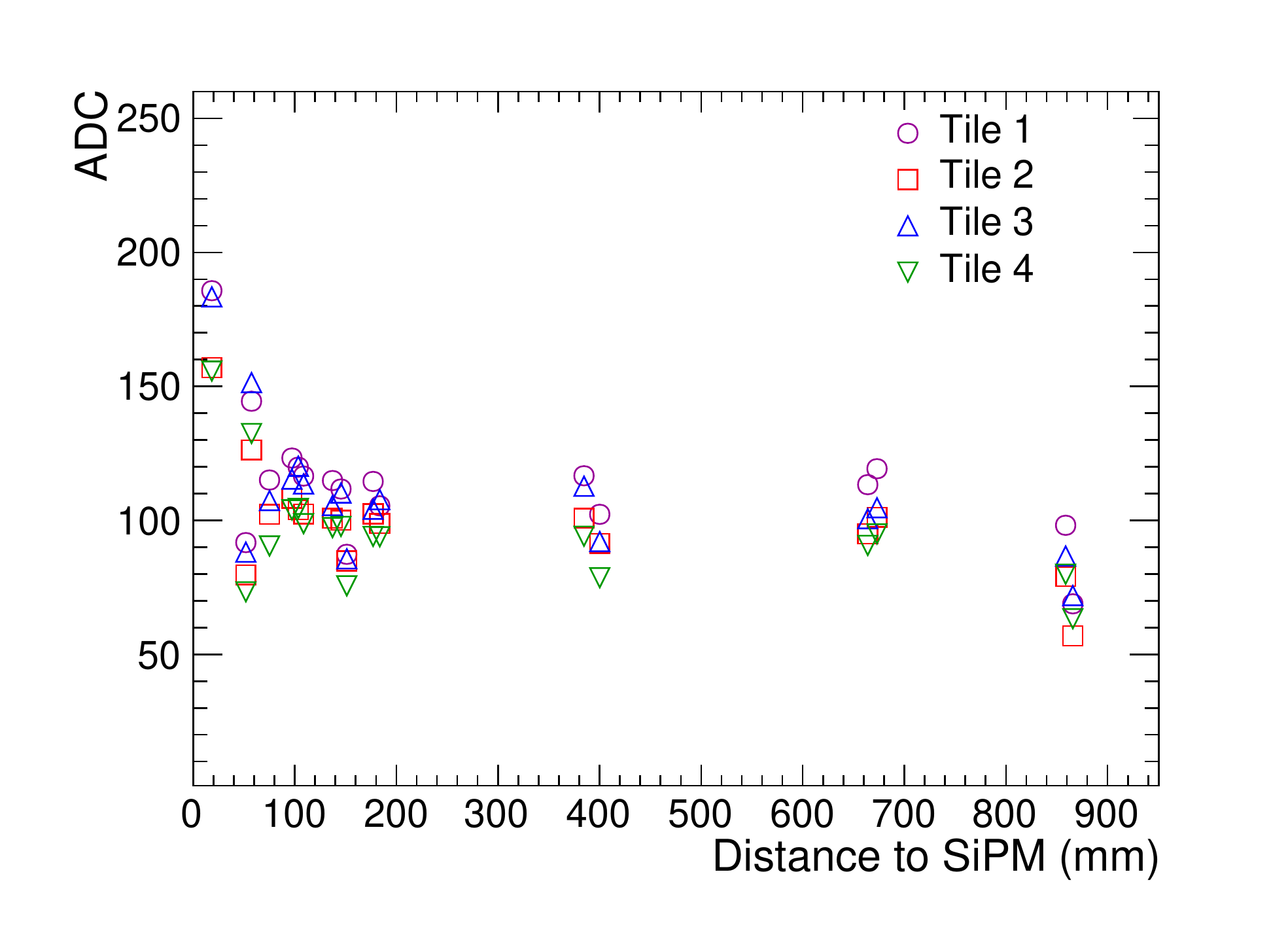}
\end{center}
\caption{Outer HCal tile scan using 16 GeV pion beam. Average ADC value in the tile plotted as a function of distance from the SiPM.
The points below 150 mm indicate an enhancement close to the SiPM.}
\label{fig:tilemapper_distance}
\end{figure}

Figure~\ref{fig:tilemapper_distance} shows the average ADC value for each scan position as
a function of the distance from the SiPM.  While the 16 GeV pion data do not
show as much of an enhancement near the SiPM as the LED scan, it can be
seen that for points less than 15~cm away from the SiPM that there is a
strong rise in the average ADC as the distance to the SiPM decreases.  This is most likely
due to the fact that some of the light in the fiber is carried in the cladding, which has
a very short attenuation length, and is therefore lost for most positions in the tile.
Studies of small double-ended scintillating tiles have indicated that up to 50\% of the light
is carried in the cladding, though this is with LED light rather than scintillation light.
Here the results indicate that about 33\% of the light is carried in the cladding.
The area in which more light is collected due to light being present in the cladding is of
order 5 cm$^2$ right around the SiPM mounting, which is at the back of the calorimeter.
The spatial density of shower particles is lowest at the back of the calorimeter and therefore
this small amount of additional light has a negligible effect on the determination of the
shower energy.

\subsection{Geometry}

\begin{table}[!hbt]
  \centering
  \caption{Design parameters for the Hadronic Calorimeter Prototype.}
  \begin{tabular}{lr}
    \hline
	Parameter & Inner/Outer HCal value \\
    \hline
 	Inner radius (envelope) & 116/182 cm \\
 	Outer radius (envelope) & 137/269 cm \\
 	Material 		& ASTM A36 Steel \\
 	Number of electronic channels (towers) & 16 \\
 	Absorber plate thickness at inner radius &  1.02/2.62 cm \\
 	Absorber plate thickness at outer radius &  1.47/4.25 cm \\
 	Total number of absorber plates & $21$ \\
 	Tilt angle (relative to radius) & 32/12 $^{\circ}$\\
 	Scintillator thickness & 0.7 cm \\
 	Gap thickness & 0.85 cm \\
 	Sampling fraction at inner radius & 0.078/0.037 \\
 	Sampling fraction at outer radius & 0.060/0.028\\
    \hline
  \end{tabular}
  \label{tab:ihcalpar}
\end{table}

\begin{figure*}[!hbt]
  \begin{center}
     \subfloat[]{\includegraphics[width=0.45\textwidth]{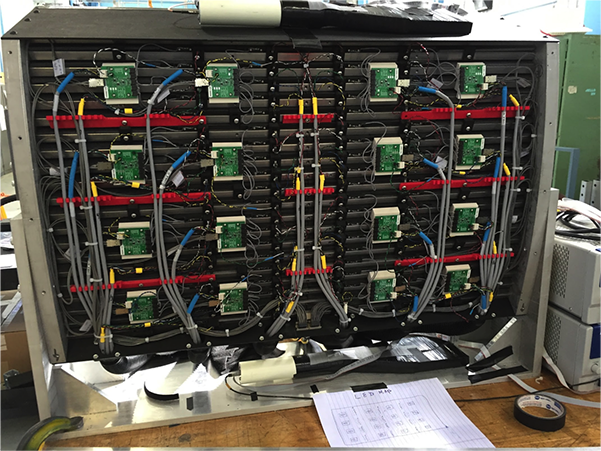}}
     \qquad
     \subfloat[]{\includegraphics[width=0.45\textwidth]{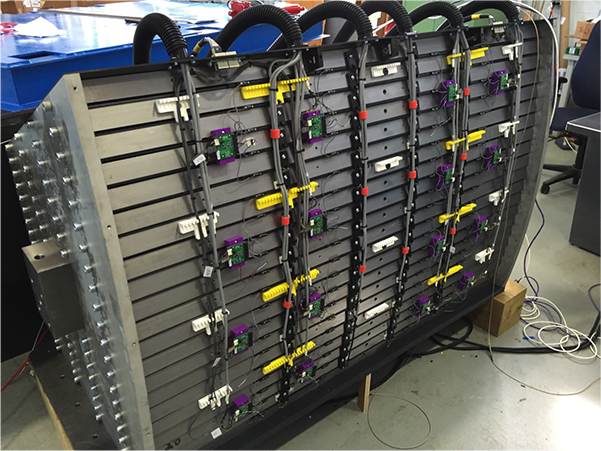}}
  \end{center}
  \caption{Fully assembled (a) inner and (b) outer HCal. Each section has 20 steel
  absorber plates stacked together and 80 scintillating tiles are inserted between
  them. SiPMs read out from five tiles are ganged together as a tower. This results
  in a total of 16 towers equipped with SiPM sensors, preamplifiers, and cables
  carrying the differential output of the preamplifiers to the digitizer system.}
  \label{fig:HCal_inner_outer}
\end{figure*}

Table~\ref{tab:ihcalpar} shows the basic mechanical parameters of the inner
and outer HCal prototypes.  The major components are 20 steel absorber plates and 80
scintillating tiles which are read out with SiPMs along the outer radius of the detector.
The SiPMs from 
five tiles are connected passively to a preamplifier channel.
This resulted in a total of 16
towers, 4 in $\phi$ by 4 in $\eta$, equipped with SiPM sensors, preamplifiers, and cables
carrying the differential output of the preamplifiers to the digitizer
system. Figure~\ref{fig:HCal_inner_outer}~(a) shows the fully assembled inner HCal. Sixteen
preamplifier boards corresponding to the 16 towers are visible. In order to make
the whole system light tight, the front and back sides were covered with electrically
conductive ABS/PVC plastic. This material quickly diverts damaging static charges if
there is a buildup. Corners were sealed with light tight black tape.
No light leaks were observed during the entire data taking period.

Since the same bias voltage is supplied to all five SiPMs in a given
tower, the SiPMs must be gain matched.  The SiPMs are sorted and
grouped to towers according to the manufacturer's measurements. The SiPM sensors,
preamplifiers, and cables are arranged on the outer radius of the inner HCal. The
interface boards mounted on the side of the modules monitor the local temperatures and
leakage currents, distribute the necessary voltages, and can provide bias corrections for
changes in temperature and leakage current.

Figure~\ref{fig:HCal_inner_outer}~(b) shows the fully assembled outer HCal. The design of
the outer HCal is similar to the inner HCal and the prototype likewise comprises 16~towers. 
However, since the absorber occupies considerably more radial space, in order to
have a minimum thickness of $5.5 \lambda_{I}$ a smaller tilt angle as noted in Table~\ref{tab:ihcalpar} is needed to preserve
the four-tile-crossing geometry.
The outer HCal SiPM sensors and electronics were arranged on the
outer face of the detector, as seen in Figure~\ref{fig:HCal_inner_outer}~(b).

\section{Readout Electronics and Data Acquisition}
\label{sec:electronics}

\subsection{Overview}

A common electronics design has been chosen for the readout of the sPHENIX EMCal and HCal
detectors using commercially available components.  The design uses SiPMs from
Hamamatsu as the optical sensors to read out the
calorimeters. Signals from the SiPMs associated with a calorimeter tower are passively
summed, amplified, shaped, and differentially driven to a digitizer system located near
the detector. The signals are continuously digitized at 60 MHz and delayed in a digital
pipeline pending a Level-1 trigger from the trigger system.  Upon receipt of a Level-1
trigger, the data for 24 time slices corresponding to the triggered event for all towers
in the EMCal and HCal are recorded. In addition to the calorimeters, signals
from the FTBF beam line Cherenkov counters (used to tag particular particle species) and
finger hodoscopes (an 8 $\times$ 8 array of 0.5~cm wide scintillators used to determine
the beam position event-by-event) are digitized and recorded. Details of the readout
electronics for the calorimeters are discussed in this section.

\subsection{Optical Sensors}

The compact nature of the EMCal and HCal detectors and the location of the EMCal and inner
HCal inside the 1.4\nobreakdash-Tesla solenoidal field of the sPHENIX experiment require that the
optical sensors be both physically small and immune to magnetic effects.  A device with
large gain is also desirable in order to reduce the demands on the performance
specifications of the front end analog electronics.  SiPMs meet these requirements.  For
both the EMCal and the HCal, the Hamamatsu S12572-33-015P
MultiPixel Photon Counter has been selected as the preferred optical sensor.
This device was chosen in order to provide a high pixel count (40K, $15~\mu m$ pixels)
for good linearity over a large dynamic range
along with high-photon detection efficiency ($\sim 25\%$).
These SiPMs were operated at voltages approximately 4~volts above the break down voltage 
in order to maintain a nominal gain of $2.3 \times 10^5$.

\subsection{Analog Front End}

\begin{figure}[!hbt]
\begin{center}
\includegraphics[width=\linewidth]{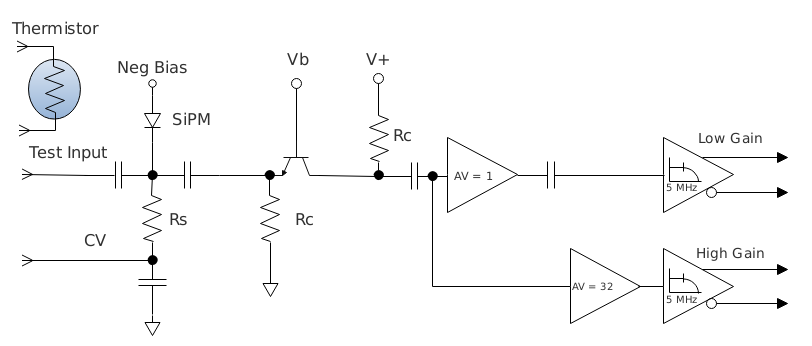}
\caption{Schematic diagram for the inner HCal preamplifier.
For the EMCal and the outer HCal preamplifier, the high gain
channel is $\times$16 the normal (low gain) channel.
In addition, the EMCal design included a programmable
selection of either the high gain or
normal gain channel as output to the digitizers.
}
\label{fig:preamp}
\end{center}
\end{figure}

The preamps used to amplify SiPM signals from both the EMCal and HCal are of the same design, 
differing only in packaging. The signals from the SiPMs
associated with a calorimeter tower are passively summed and then amplified. The amplifier
front end is a common-base configuration acting like a transresistance amplifier or
``current conveyor.'' This configuration presents a very low impedance to the SiPMs
thereby minimizing any voltage swing on the device. A charge injection circuit is included
to generate a fixed test pulse to the amplifier. The signal then passes through gain
circuitry which can select either normal gain or a high gain range, 16x normal for the
EMCal and outer HCal and 32x normal for the inner HCal.  For the EMCal, the gain range is
selectable via slow control, while for the HCal both ranges were recorded during normal
data taking.  The amplified signal is then shaped with a peaking time of 30~ns for 60~MHz
sampling and driven differentially to the external ADC electronics.
Figure~\ref{fig:preamp} shows a schematic of the preamp design.

The EMCal preamplifier module consists of eight preamp channels with 4 SiPMs per channel
laid out to match the tower geometry of a 1~$\times$~8 EMCal block. A small surface\nobreakdash-mount
thermistor is centered between the 4 SiPMs in a tower to monitor the local temperature of
SiPMs.  Located on the edge of the preamplifier module centered between pairs of EMCal
towers are 4 LEDs that are used for monitoring and testing.  The SiPMS for the HCal are
mounted on small daughter boards that are directly attached to the tiles forming an HCal
tower. The five SiPM daughter boards for a tower are connected
to an HCal preamp board located on the
detector with a short shielded twisted pair cable. An LED is also positioned near each
SiPM for monitoring and testing. A thermistor is located near one of the SiPMs in each
tower to monitor the local ambient temperature.

\subsection{Slow Control}

A slow control system is necessary to provide stable and controllable operational
parameters for the SiPMs. The sPHENIX slow control is organized in a tree structure to
facilitate the monitoring and control of a large number of channels in the final detector
configuration.  The slow control system provides: SiPM bias control, SiPM leakage current
readback, SiPM temperature measurement, input voltage and voltage regulator temperature,
pulse control for both charge injection and LED test pulse, and SiPM temperature and
leakage current compensation.  The slow control system comprises an interface board
and controller board. The interface board, located on the detector, contains ADCs for
monitoring temperatures, voltages, and leakage current, and digital-to-analog converters for adjusting the SiPM
bias voltage. In addition, the interface board provides power and bias distribution to the
preamp boards. For both the EMCal and the HCal, the interface board functionality is the same;
however, the packaging is different to account for differences in geometrical constraints
of the two systems. The interface board is connected via a bi-directional serial link to a
controller board in a nearby crate. The controller board transmits to the interface board
the parameters for gain control, temperature compensation, LED enables, and pulse
triggers, and reads back monitoring information.  Communication with the controller board
is ethernet based.

The controller boards are 6U Ethernet Telnet servers housed in external racks that control
the interface boards via isolated RS-485 serial lines over standard CAT5 cables. In
addition to providing a monitor and control portal, the controller processes temperature
and leakage current values to provide individual SiPM temperature and leakage current
compensation if so commanded.

\begin{figure}[!hbt]
\begin{center}
\includegraphics[trim={1cm 2.5cm 1cm 3cm},clip,width=\linewidth]{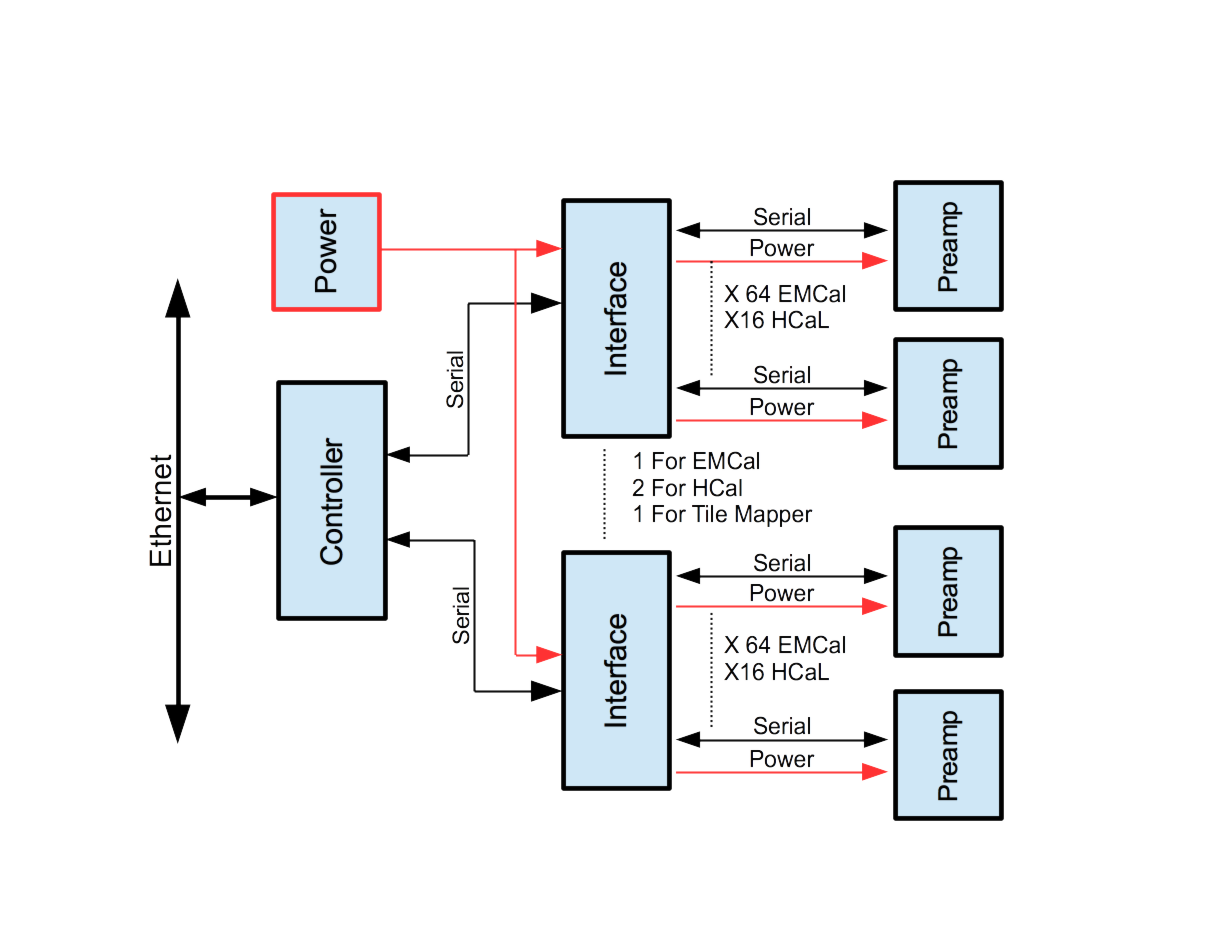}
\caption{\label{fig:slowcontrol} Schematic overview of the slow control system for the
EMCal and HCal. The interface board, connected to the preamp modules, provides
environmental monitoring along with power distribution. The controller board receives
monitoring data from the interface board and transmits to the interface board slow
control commands.  Communication with the controller board is ethernet based.}
\end{center}
\end{figure}

Two controllers are used for the T-1044 Beam Test, one for the EMCal and one for the HCal. A
single EMCal interface board serviced the 64 channel EMCal detector. A separate HCal
interface board is used for each of the 16 channel Inner and Outer detectors, and the
tile mapper. A schematic  overview of the slow control system is shown in
Figure~\ref{fig:slowcontrol}.

\subsection{SiPM temperature compensation}

\begin{figure*}[!hbt]
  \subfloat[]{ \includegraphics[width=.5\textwidth]{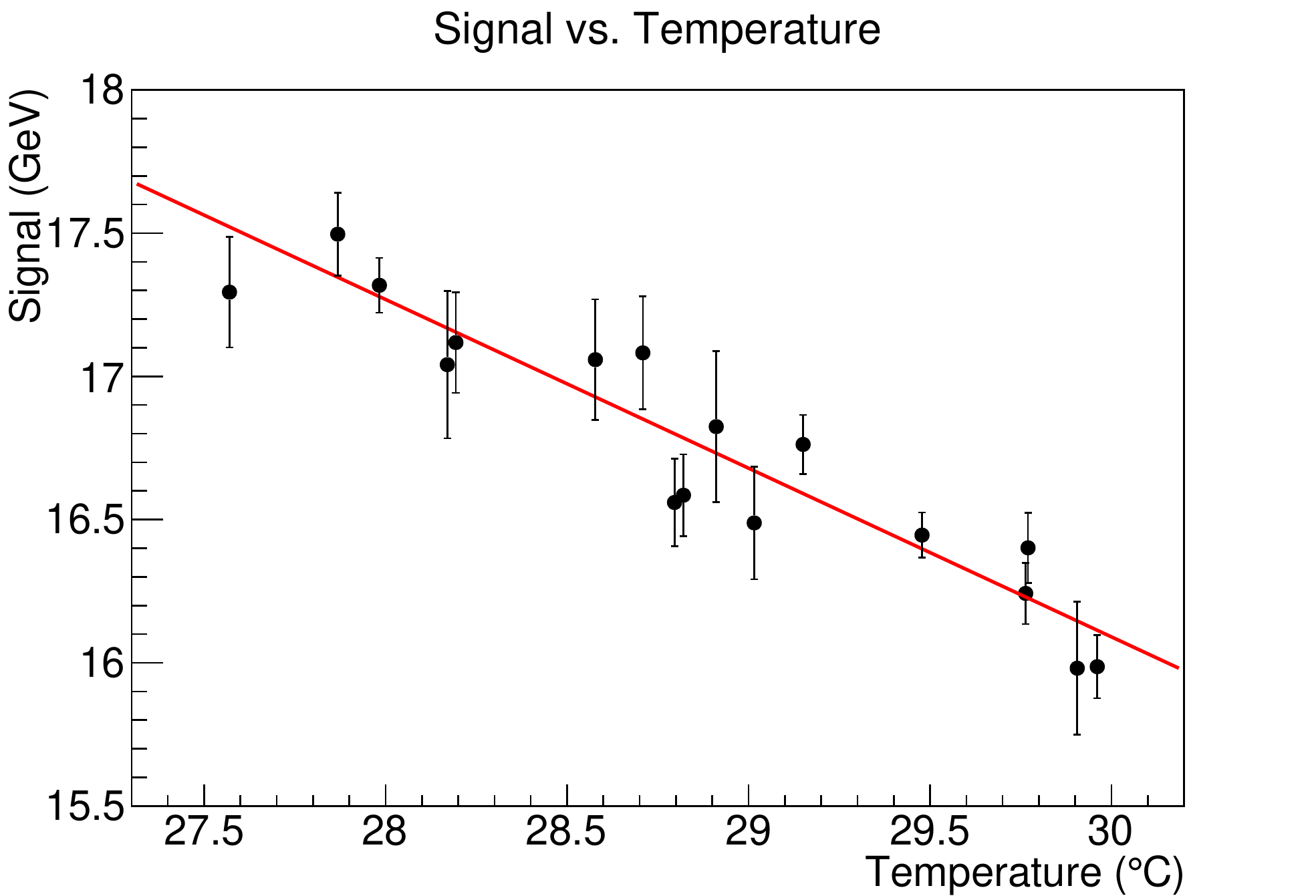}}
  \subfloat[]{ \includegraphics[width=0.5\textwidth]{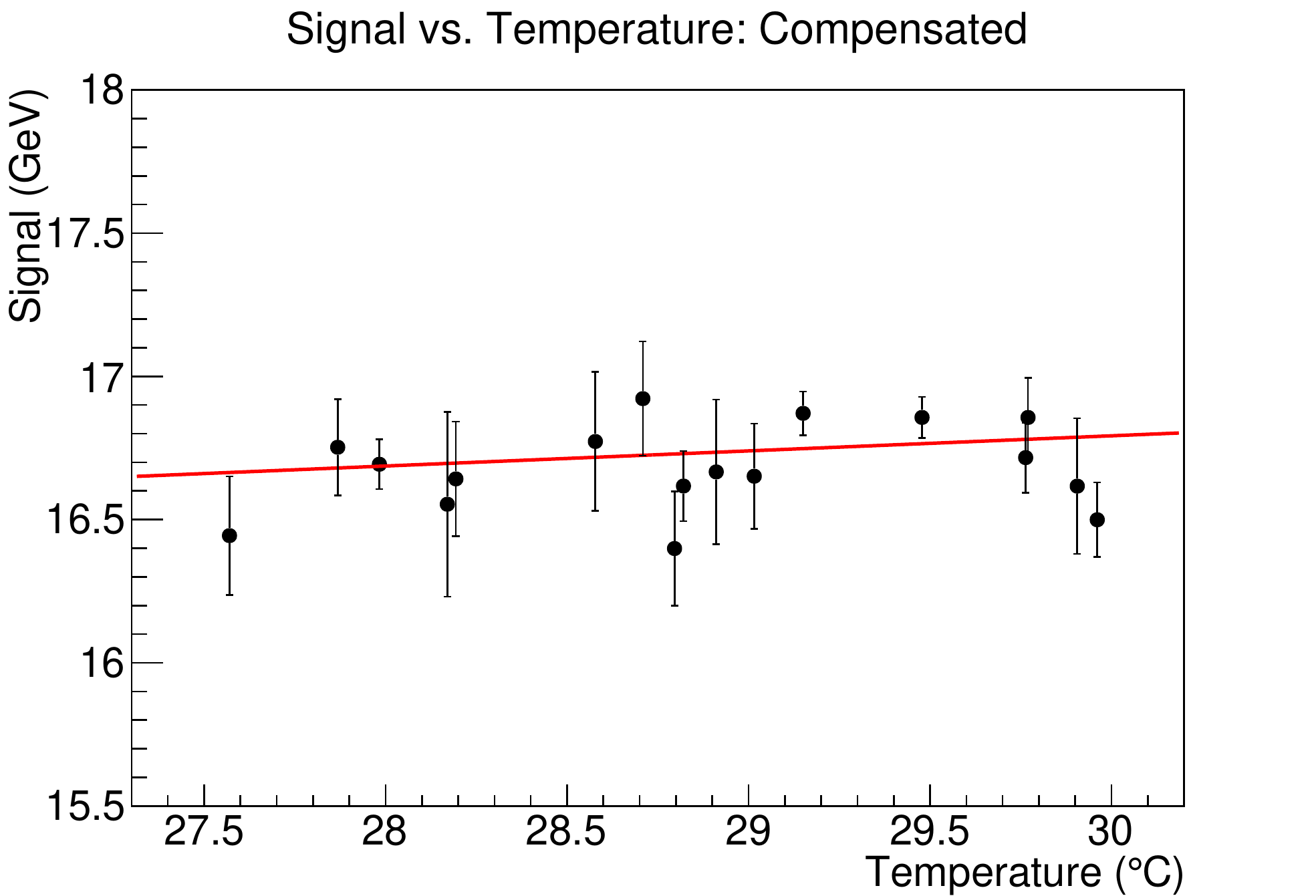}}
\caption{\label{fig:fig_tempcomp} EMCal SiPM signal vs. temperature without
(panel (a)) and with (panel (b)) temperature compensation.}
\end{figure*}

The gain of an SiPM is sensitive to the temperature.  Large variations in gain due to temperature 
variations can potentially lead to significant uncertainties in determining the energy deposited 
in the calorimeter. 
There are several approaches to stabilizing the temperature of the SiPMs or limit
the potential temperature variations of the SiPMs. The first approach is to use active cooling to 
stabilize the temperature to a known temperature using either a Peltier or liquid coolant based system.  
Another approach is stabilize the temperature using air cooling and  to perform corrections to
the data to compensate for the variations in gain as a function of temperature.  This is the 
approach that was chosen for the prototype detector. The temperature
compensation can be accomplished in one of two ways.  Firstly, the slow control system is
capable of making online gain adjustments (by adjusting the SiPM bias voltage) based on
the temperatures measured by the thermistors in the front end electronics.  Secondly, it
is possible to perform an offline correction by determining the temperature dependence of
the gain using the recorded data.  Temperature data at the SiPMs from the thermistors in
the front end electronics are recorded in the data stream.  In order to account for data
taken at different temperatures, special data was taken to calibrate the temperature
correction.  Figure~\ref{fig:fig_tempcomp} shows the calibrated signal (in GeV) vs. the
average (event-by-event) temperature across the 5~$\times$~5 region in the EMCal around a
particular tower (note that the temperature may be offset by some constant due to lack of
an absolute calibration).  Using the calibration data, an offline temperature correction
is determined and applied to correct the data.  Figure~\ref{fig:fig_tempcomp} (a) shows
example data with no temperature compensation applied, which shows a variation of
(-3.68~$\pm$~0.29)\%/$^\circ$C.  Figure~\ref{fig:fig_tempcomp} (b) shows example data with
the offline temperature compensation applied.
The slope is (0.33~$\pm$~0.30)\%/$^\circ$C,
which is effectively consistent with no temperature
dependence within the statistical uncertainty.

\subsection{LED Monitoring}

To provide monitoring and testing of the calorimeter electronics, an LED pulser system is
used. For the EMCal, four 405 nm LEDs are placed on the preamp board such that they are
centered between 4 EMCal towers, 2 towers associated with the preamp board and 2 towers
associated with the neighboring preamp board.  For the HCal, a 405 nm LED is embedded in
the edge of each of the five tiles associated with an HCal tower. The LEDs are driven
with a fixed amplitude pulse and can be individually pulsed using a programmable
driver circuit through the slow control system.

\subsection{Digitizers}

The analog signals from the front end amplifiers are transmitted differentially over a
10~m Hard Metrics 16 channel signal cable to a custom digitizing system originally
developed for PHENIX~\cite{Anderson:2011}.
The signals are received differentially and digitized by a 12 bit flash ADC
running at a 60 MHz sampling frequency. The output of the ADC is transmitted to a local
FPGA which provides a 4~$\mu$s pipeline delay for buffering events for a Level-1 trigger.
Upon receipt of a Level-1 trigger, the ADC data for the 24 time samples for the triggered event is transmitted via
optical fiber to a PHENIX Data Collection Module (DCM).  The formatted data is transmitted
to a local computer for logging to disk. The system is designed to operate at the planned
sPHENIX maximum Level-1 trigger acceptance rate of 15 kHz.

\subsection{Data Acquisition System}

The current data acquisition system in use for most sPHENIX R\&D-level efforts is called
Really Cool Data Acquisition (RCDAQ)~\cite{RCDAQ}.  
RCDAQ is client-server based, and can be controlled from multiple
clients. There is no ``central console'' for the operation of the Data Acquisition (DAQ)  system. This allows,
for example, the DAQ to be started from the beam enclosure while an access is underway to
verify the proper state of all components before ending the access.  RCDAQ offers an
online monitoring stream, which provides the most recent data on a best-effort basis (the
online monitoring is not allowed to raise the DAQ busy and throttle the data rate). This
monitoring allows any tripped voltage supplies, noisy, or dead channels to be recognized
in a timely manner.  Additionally, RCDAQ allows the capture of any kind of ancillary data
that can be accessed from the DAQ computer, such as temperature readings, voltage levels,
and camera pictures. These are embedded in the primary data stream and cannot be separated
from the data, and therefore cannot get lost.  This additional information provides the
ability to perform a ``forensics-type'' investigation if there is a problematic result or
if one finds confusing or incomplete logbook entries.

\section{Test Beam}
\label{sec:testbeam}

\begin{figure}[!hbt]
\begin{center}
\includegraphics[width=\linewidth]{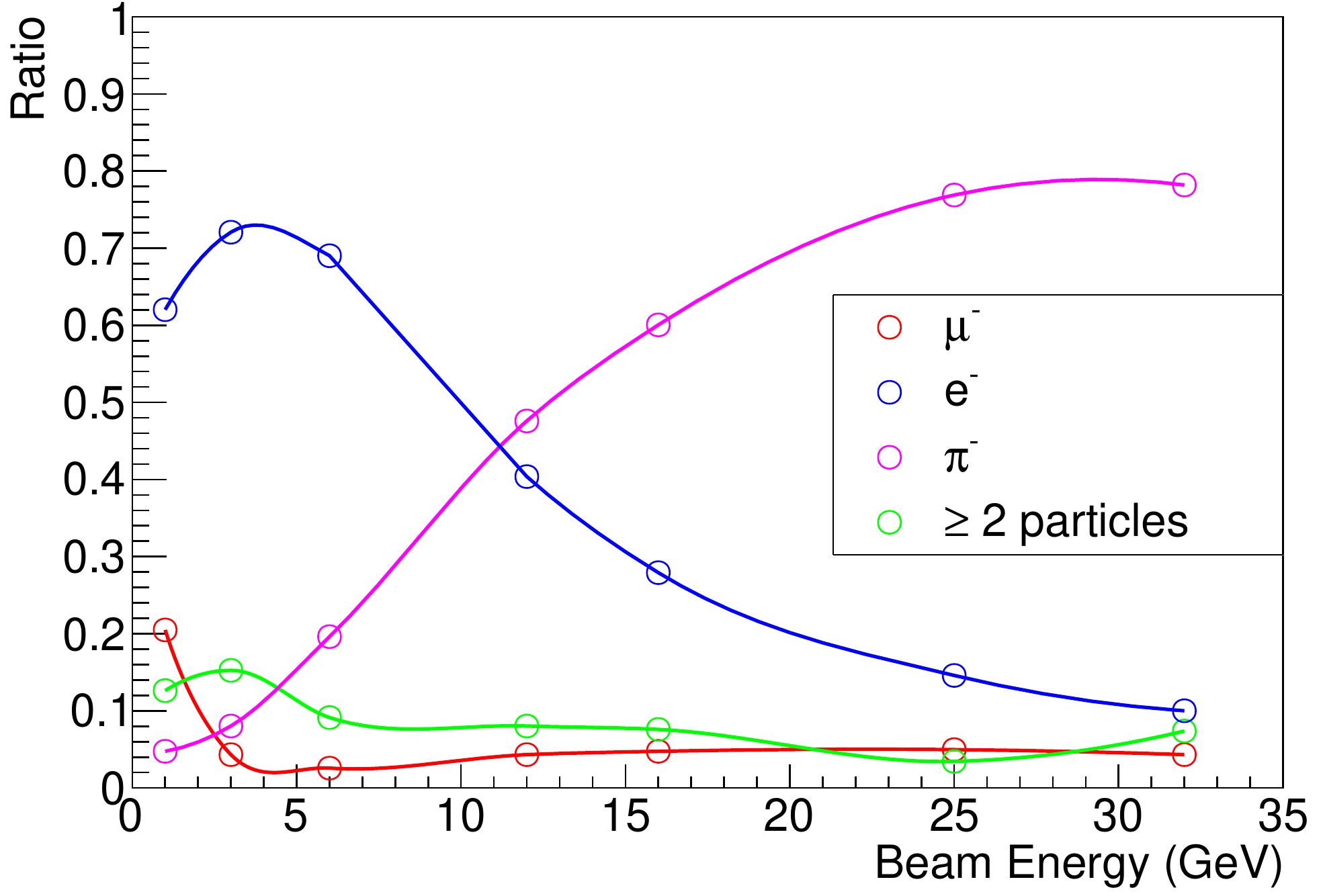}
\end{center}
\caption{The relative abundance of particle species present in
the FTBF negatively charged beam as a function of beam energy~\cite{ftbf}.}
\label{fig:beamprofile}
\end{figure}

Testing of the prototype detectors was performed at the Fermilab Test Beam Facility (FTBF)
designated as the T-1044 experiment. The facility has two beamlines which can produce a
variety of particle types over a range of energies up to 120~GeV. The T-1044 experiment
used the MTest beamline which has two modes of operation; primary protons at 120 GeV and a
secondary mixed beam consisting primarily of pions, electrons and muons with energies
ranging from 1 to 60 GeV of either positive or negative charge. The beam energies used for
T-1044 were secondary beams of 1, 2, 4, 6, 8, 12, 16, 20, 24, and 32 GeV, and primary
protons at 120 GeV. The beam is delivered as a slow spill with a 4 sec duration once per
minute with a maximum intensity of approximately $10^5$ particles per spill. 
The momentum spread of the beam at the FTBF depends on the beam energy, beam tuning parameters and collimator settings. For our measurements of the calorimeter resolution, these parameters were set to provide a momentum spread of $\approx 2\% ~\Delta p/p$ over the energy range from 2-16~GeV, which is consistent with the value estimated by simulation calculations of the beam line~\cite{meson_mom}, and by our own measurements of the beam with a lead glass detector and by other test beam experiments~\cite{Tsai:2012cpa}.
The beam
spot size is also dependent on the beam energy and tune and ranges from approximately
$0.6$~cm to several centimeters in size.  The secondary beam composition is plotted in
Figure~\ref{fig:beamprofile} showing the relative contribution of electrons, muons and
hadrons present in the beam as a function of energy~\cite{ftbf, Feege:2011dsa}. The kaon
content in the beam is expected to be around 1\% for beam energies between
20-32~GeV~\cite{Blatnik:2015bka}.

The FTBF also provides a number of detectors for test beam groups to use. These include two
differential gaseous Cherenkov counters upstream of the MTest enclosures, a lead glass
calorimeter, multi-wire proportional chambers (MWPC) and trigger counters~\cite{ftbf}.

The Cherenkov counters are used for offline
discrimination between pions and electrons on an event-by-event basis.  The gas pressures
of the counters are set between the pion and electron thresholds. The inefficiency of the
Cherenkov counters resulted in only a few percent contamination for the pion samples.

\begin{figure}[!hbt]
\centering
\includegraphics[width=\linewidth]{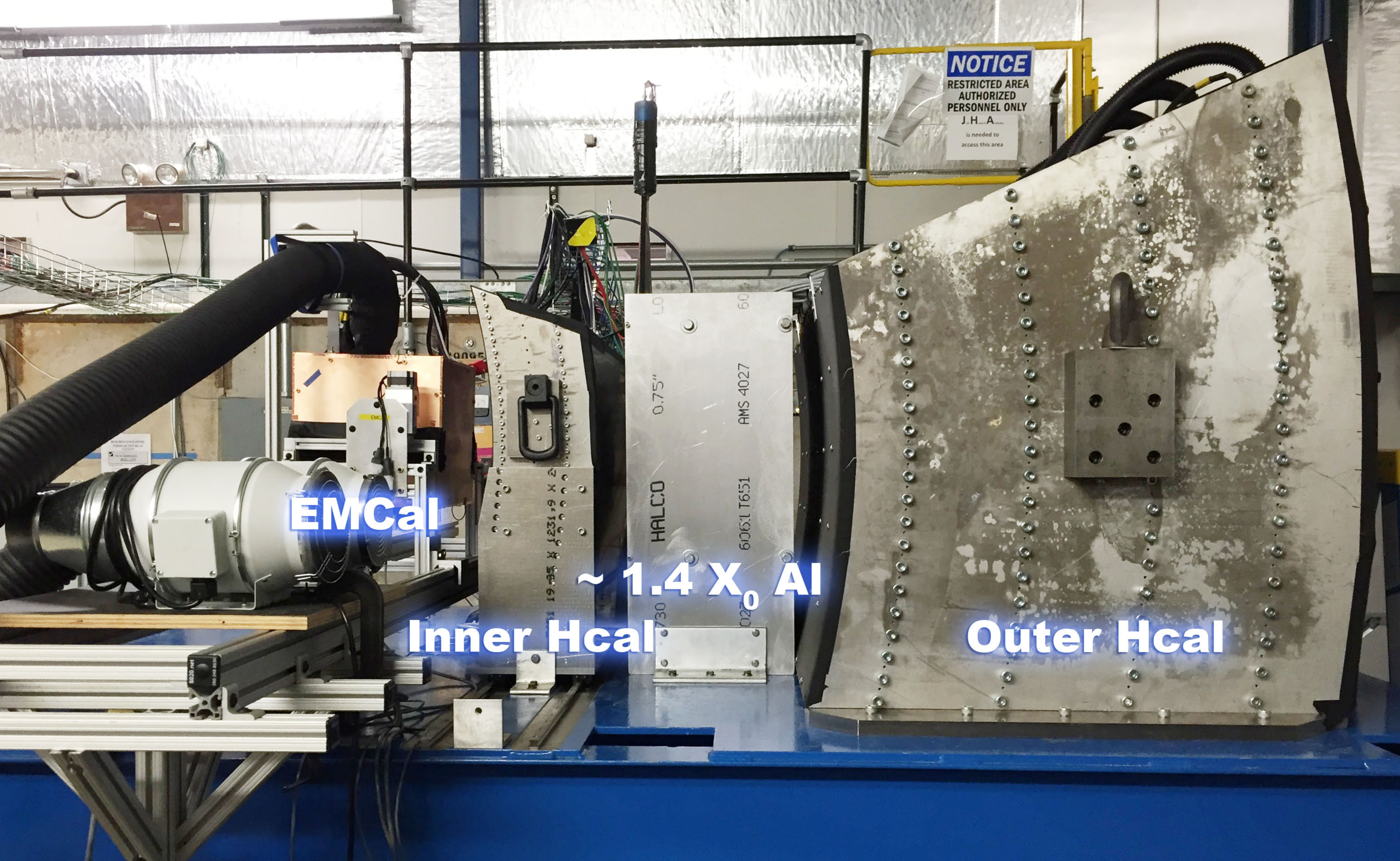}
\caption{T-1044 test beam setup is shown where the beam enters the detectors from the left
of the image. The EMCal, Inner HCal, mock cryostat of 1.4 $X_{0}$ Al and the Outer HCal
are all labeled.}
\label{fig:testbeam_setup}
\end{figure}

This experiment used the MT6.2C and MT6.2D areas of the MTest beam line~\cite{ftbf}. For
the initial tests of the EMCal, the EMCal detector was placed on the MT6.2C motion table.
The motion table allowed the detector to be moved with respect to the beam remotely.
The EMCal detector is rotated horizontally by an angle of
10 degrees with respect to the beam axis to prevent channeling effects. 
Additionally, it was also studied at an angle of 45 degrees.
For the second half of running, the EMCal was moved to be directly in front of the Inner
HCal allowing for combined EMCal/HCal testing.  Figure~\ref{fig:testbeam_setup} shows a
picture of the prototype setup in the MT6.2D area.

A hodoscope and veto counters were installed upstream of the EMCal to allow for the
selection of beam particles impinging on the EMCal detector.  The hodoscope was provided courtesy of the UCLA group~\cite{Tsai:2012cpa,Tsai:2015bna} and consisted of
16 0.5~cm finger counters (eight vertical and eight horizontal) read out with SiPMs.  The signals
from the SiPMs were amplified, shaped, and read out using the digitizers. Four
scintillator veto counters surrounded the hodoscope and are read out using PMTs and digitized
using the digitizers. If any of the veto counters measures
energy above a certain threshold, the event is rejected due to the position of the beam.
As shown in
Figure~\ref{fig:beamprofile}, a small fraction of events have more than one particle in one
event. In order to remove those events, each event is required to have only one valid
horizontal and vertical hit in the hodoscope.

\begin{figure*}[!hbt]
\centering
\includegraphics[width=.75\linewidth]{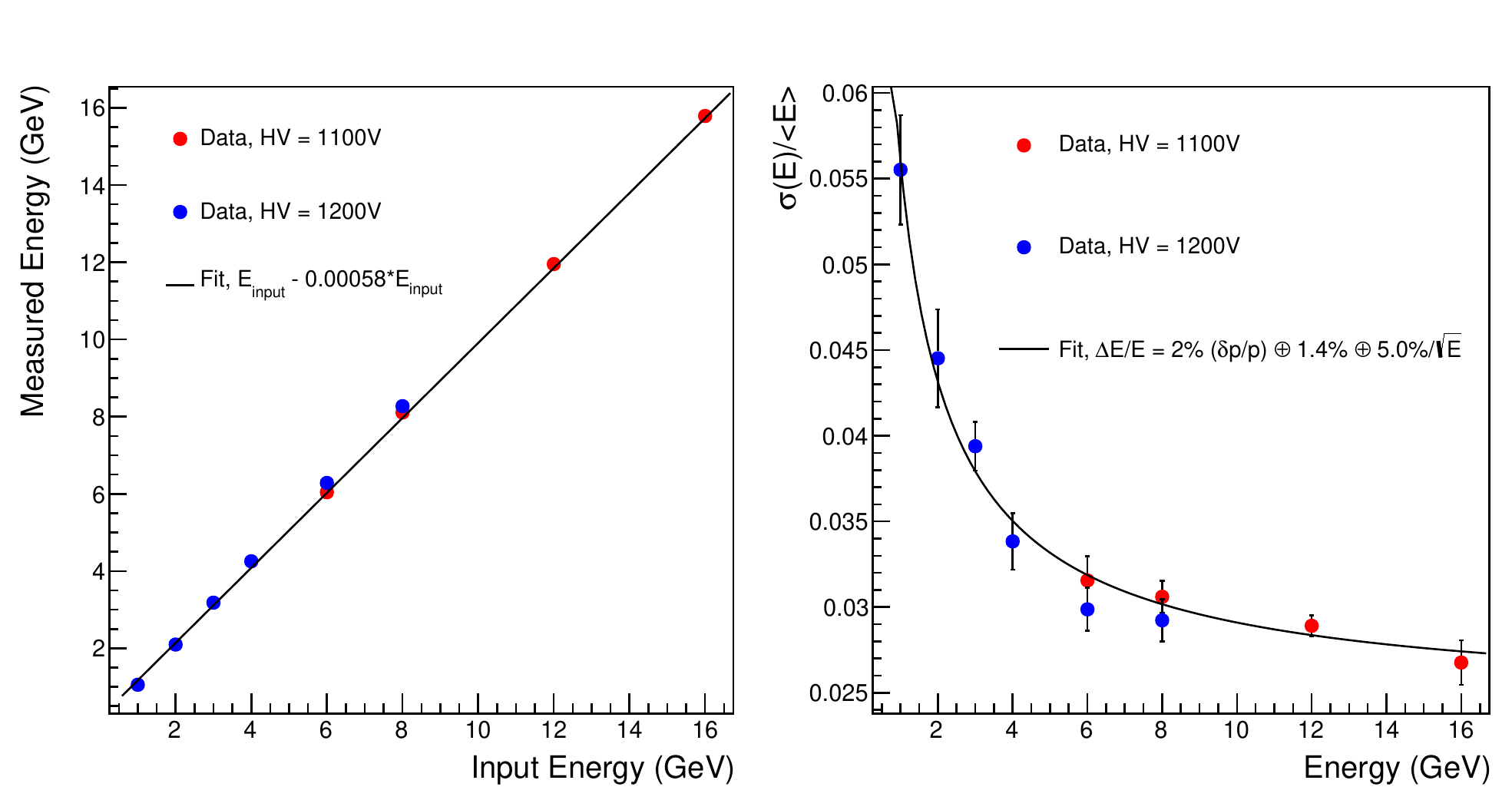}
\caption{Pb Glass linearity (left) and energy resolution (right) measured with two high
voltage settings of 1100V (red) and 1200V (blue).}
\label{fig:Pbglresolution}
\end{figure*}

A $45 \times 15 \times 15$~cm$^3$ SF-5 Pb-glass calorimeter is used to double check the
test beam energy scale and momentum spread. 
Pb-glass of this type is known to have a resolution of
$(5.6\pm 0.2)\%/\sqrt{E}$~\cite{Brown:1986fb}.  The energy
resolution was measured at two different operating voltages 1100V and 1200V and obtained the result of
$2.0\%(\delta p/p)~\oplus~1.4\%~\oplus~5.0\%/\sqrt{E}$, as shown in
Figure~\ref{fig:Pbglresolution}.

\section{Simulations}
\label{sec:simulations}

\begin{figure}[htb]
\centering
\includegraphics[width=1.0\linewidth]{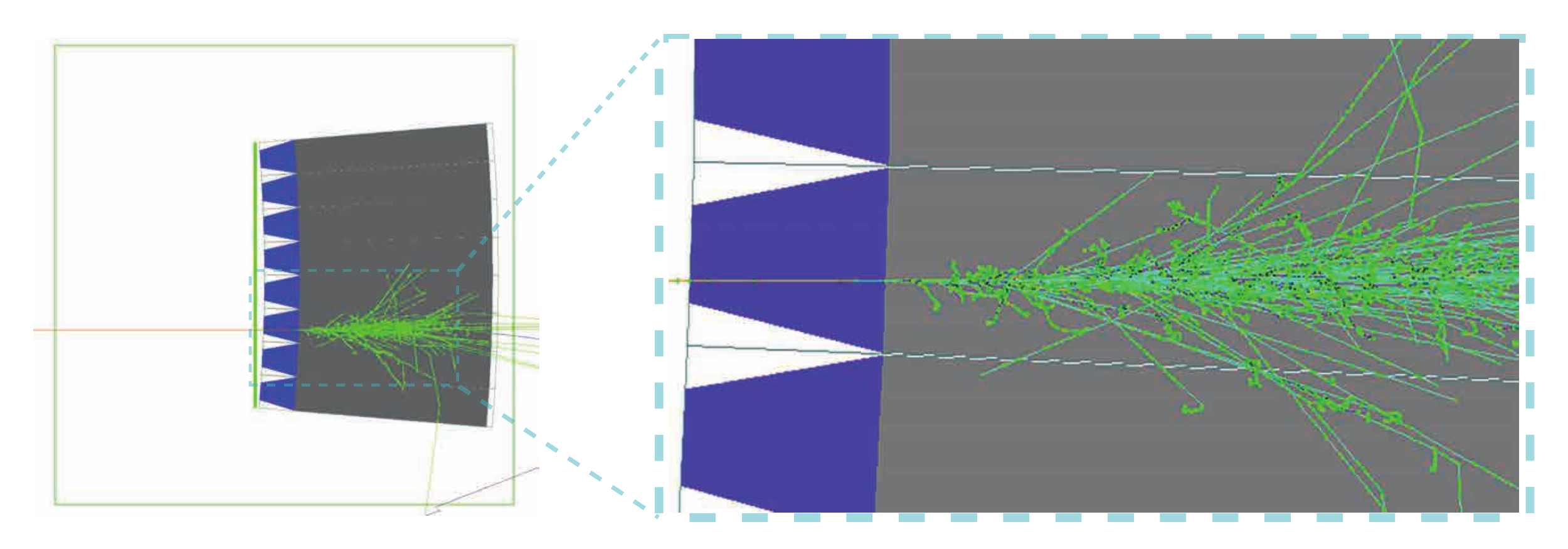}
\caption{Side view of an 8~GeV electron shower overlaid with the whole EMCal prototype (left) and zoom-in (right). The incoming electron (red line entering from left) passes through the light guide (blue
trapezoid) and develops an EM shower in the W/SciFi EMCal blocks
(gray blocks). A 1/10-inch-thick G10 sheet (green vertical plate) is placed
before the EMCal to represent the average thickness of the electronics and cooling assembly. Only particles with an energy higher than the critical energy in tungsten (7.8 MeV) are shown.}
\label{Fig_Sim_EMShower}
\end{figure}

\begin{figure}[htb]
\centering
\includegraphics[width=1.0\linewidth]{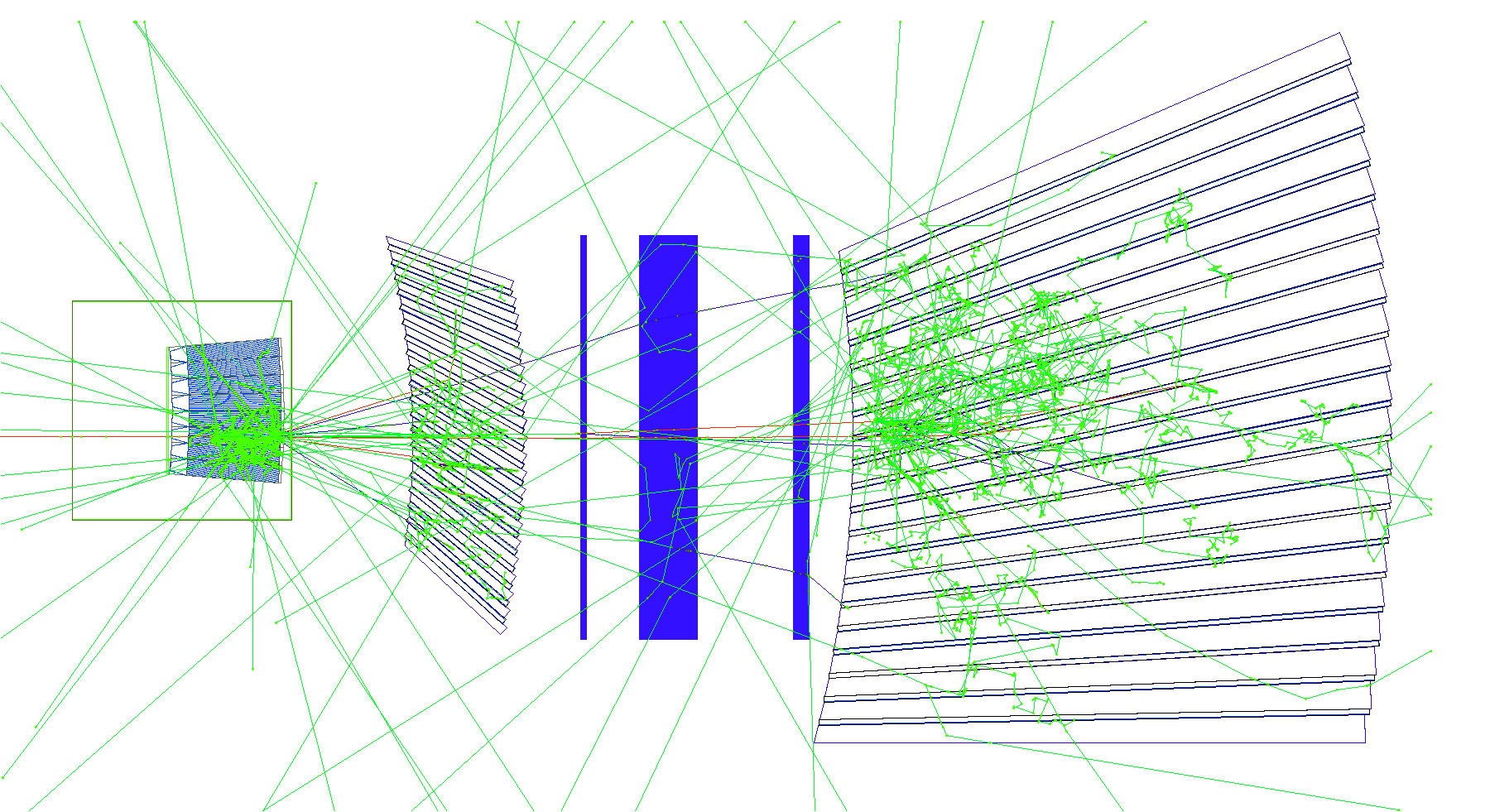}
\caption{Side view of a 32~GeV $\pi^-$ shower as simulated in the EMCal and HCal prototype.
The incoming pion (red line on the left edge) starts to develop a shower in the EMCal
(left box), which is further absorbed in the inner HCal (tilted plates in the middle),
three aluminum plates as a mock up of the sPHENIX magnet (blue block) and the outer HCal
(tilted plates on the right side).}
\label{Fig_Sim_PionShower}
\end{figure}

At this stage of the detector construction the simulation is still under development and subject to validation based on experimental measurements. The sample of results shown in this paper illustrates the methodological approach adopted to achieve a realistic model of the detector performance. 
The simulation reported here were performed using the pre-defined \mbox{\sc QGSP\_BERT\_HP} 
physics configuration distributed in \geant version 4.10.02-patch-01~\cite{Agostinelli:2002hh, Allison:2006ve}. Based on the requirements of the sPHENIX experiment and the ongoing validation process, customized physics configurations can be further investigated.

Both the EMCal and the HCal have been simulated using 
these simulation settings
.
Their location and internal structure are carefully assigned in order to match that of the
actual test beam device.
Figure~\ref{Fig_Sim_EMShower} shows typical \geant events of 8~GeV electron showers in
the EMCal. Most of the shower particles are contained within the EMCal, and the core of
the shower is sampled by multiple rows of fibers. Figure~\ref{Fig_Sim_PionShower} shows a
30~GeV~$\pi^-$ shower in both the EMCal and the HCal prototypes, in which the shower initiates
in the EMCal and most shower particles are absorbed by the three segments  of calorimeters.

After the \geant simulation stage, digitization was implemented with the following four steps:
\begin{enumerate}

\item Energy depositions for each \geant tracklet in the scintillation volume are
      collected in the sPHENIX analysis framework.

\item The Birks' law of scintillator non-linearity~\cite{Birks:1951boa} with  
      an ansatz
      Birks' constant of $k_B$~=~0.0794~mm/MeV~\cite{Hirschberg:1992xd} is implemented to
      convert ionizing energy deposition to visible energy that is proportional to the
      expected number of photons produced in the scintillator. 
      The final selection of Birks' constant to be used in sPHENIX simulation
      still subject to further optimization.

\item The visible energy in each calorimeter tower is summed in a timing window of 0-60~ns
      to calculate the mean number of active pixels in the SiPM readout. In the case of
      the EMCal, the sum of visible energy is also modulated by the position of
      scintillation light production, which accounts for the measured attenuation in the
      scintillation fiber and the non-uniformity in the light collection efficiency for
      the light guide as shown in Figure~\ref{fig:EMCal_LG_REL_EFFICIENCY}.  The scale of
      the mean number of active pixels is set by the mean active pixel count as measured
      in cosmic tests of the EMCal and HCal. The actual active pixel number is a random
      number following a Poisson distribution with a parameter of the mean number of
      active pixels.

\item In the last step, the ADC for each readout channel is proportional to the sum of the
      actual active pixel number and a random number following the pedestal distribution.
      The sum is scaled to an ADC value using measured pixel/ADC value from cosmic tests
      and discretized to integer ADC value.

\end{enumerate}

The sPHENIX simulations have been integrated with the sPHENIX software
framework~\cite{sPHENIXSoftwareFramework}, enabling the same  analysis software setup to be used to
analyze both the simulated and the beam test data.
The simulated data are compared with the real data
as discussed in Section~\ref{sec:results}.

\section{Results}
\label{sec:results}

The data from the test beam T-1044 at Fermilab are studied for three different
configurations. First, the EMCal is tested as a standalone calorimeter, then the inner and
outer HCal are tested, and finally all three calorimeters are tested together as a calorimeter system. 
For each of these combinations, energy linearity and resolution are measured at selected beam energies. The resolution data points, $\sigma(E)/\left< E \right>$, are then fit with empirical parametrization functions of the beam energy, $E$, which are quadratic sums of constant and statistical terms, $\sigma(E)/\left< E \right> = \sqrt{a^2 + b^2/E}$. 
For each fit, the $p$-value for the hypothesis of the fit function is calculated using the $\chi^2$ goodness-of-fit test, assuming the goodness-of-fit statistics follows a $\chi^2$ p.d.f. with the degree of freedom as the number of data points minus two (the number of free fit parameters)~\cite{Tanabashi:2018oca}.
We consider a fit with $p$-value $> 0.05$ would indicate the data follow the empirical parametrization within its statistical uncertainty. 
Meanwhile, we do not expect these curves to describe linearity and resolution with high precision, while the key point for the comparisons and parameterizations is to demonstrate we can describe the general trends of these data.
Therefore, for the case where $p$-value$< 0.05$, we would conclude the data suggest additional features of energy dependence than these simple parametrizations. However, the parametrization is still valid for qualitatively comparing to the performance specifications of sPHENIX and to other calorimeter systems parameterized in the same way. 

\subsection{EMCal Calibration and Check}

\begin{figure}[!hbt]
  \begin{center}
  \includegraphics[width=\linewidth]{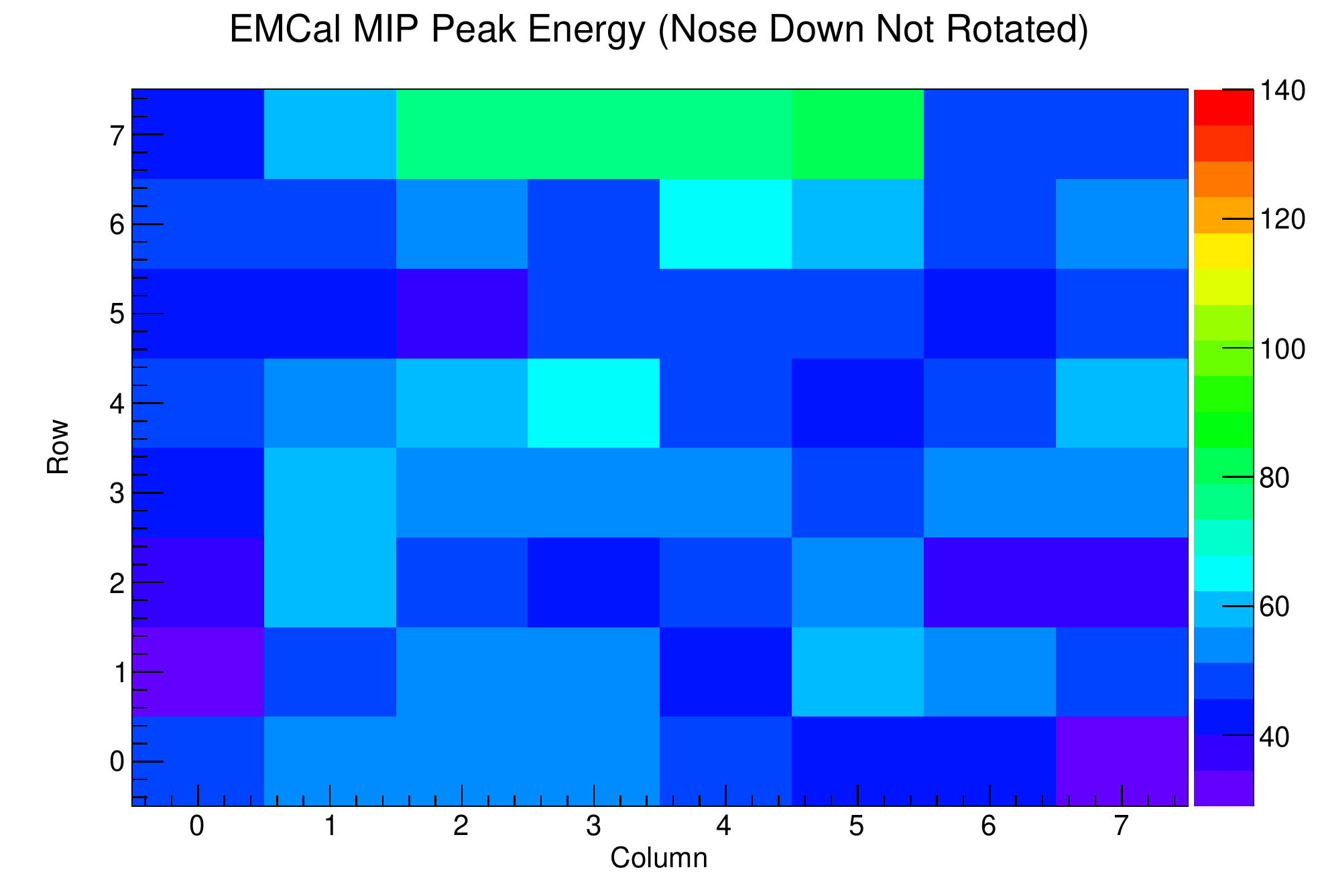}
 \end{center}
  \caption{EMCal MIP peak ADC for each EMCal tower index in columns and rows, used for energy calibration. See text for details.}
  \label{fig:horz_hodo}
\end{figure}

The relative variation of energy response for EMCal towers is calibrated using minimum ionizing particle (MIP) calibration runs.
In these data, the EMCal is rotated downward from its nominal position with the beam passing perpendicular to the EMCal towers.
The beam is centered on one column per run, such that the beam passes through all 8 towers in a column.
In order to avoid events in which a proton has initiated hadronic showers in the EMCal,
the MIP calibration events are selected by requiring signals above the
pedestal for each tower in the column of interest and no counts above the pedestal for all
other towers. The ADC spectrum for each tower in the column of interest is fit with a Gaussian+Laundau function.
The MIP peak ADC value is extracted from each fit as the ADC value corresponding to the maximum point of the fit curve.
After repeating this study for all eight columns, the MIP peak ADC for all towers are collected as shown in Figure~\ref{fig:horz_hodo}.
Four towers in the middle of row-7 show higher response in the MIP amplitude. This observation was confirmed by sending the beam from 
row-0 to row-7 and in the reverse direction for each column of towers.
The calibrated energy as shown in Section~\ref{sec:EMCal_Measurements} is corrected for the relative variation of energy response for EMCal towers by dividing the raw ADC of each tower with its MIP peak ADC prior the sum of the tower-cluster energy.

\begin{figure}[!hbt]
  \begin{center}
   \includegraphics[width=\linewidth,trim={2cm 0 5cm 0},clip]{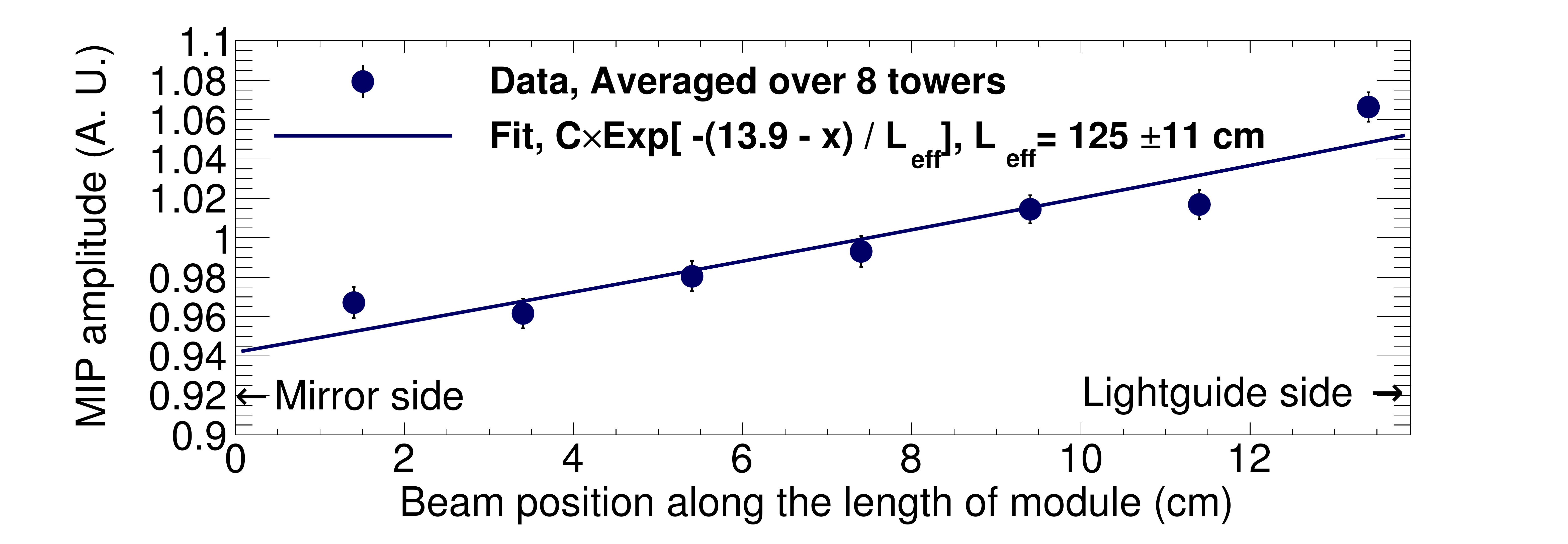}
 \end{center}
  \caption{EMCal MIP peak amplitude plotted against the beam position along the length of module. The data points, which are the average response from eight towers, are fit with an exponential function (curve).}
  \label{fig:attenuation}
\end{figure}

In order to quantify the attenuation of scintillation light inside the EMCal blocks, a 120~GeV proton beam is scanned along the longitudinal direction of the towers in the same setup as the above MIP calibration. The beam traverses the EMCal perpendicular to the length of the block from the non-tapered side of the EMCal block. The MIP signal amplitude is plotted versus the beam position along the block length dimension as in Figure~\ref{fig:attenuation}.
The data points are fit with an exponential function to extract the effective attenuation length of scintillation light in the block, $L_{\rm eff}$. 
The result is $L_{\rm eff} = 125\pm11$~cm, which is much longer than the length of the block.
This longitudinal position dependent scintillation light yield is used in the simulation in order to describe scintillation light propagation inside the EMCal.

An upper limit on the Cherenkov background, produced when charged particles pass through
the acrylic light guides, is estimated using dedicated runs in the test beam.  With the
EMCal towers rotated perpendicular to the incoming beam, 120~GeV protons are set to pass
through a column of light guides or a column of SPACAL towers. Events in which a proton
initiates a hadronic shower are rejected by vetoing events with non-zero energy in the
EMCal towers other than the column being studied. The mean energy from the Cherenkov
background when the proton beam passes through the light guide is found to be less than 11\% (90\%~C.L.) of the MIP energy when a proton passes perpendicularly through the EMCal
towers.
Since in the nominal orientation of the EMCal,
the incoming particle travels away from the photon sensor, the Cherenkov background
during physics data taking is expected be significantly smaller than this estimated upper
limit.

\subsection{EMCal Measurements}
\label{sec:EMCal_Measurements}

\begin{figure*}[htb!]
  \centering
  \includegraphics[width=1.0\textwidth]{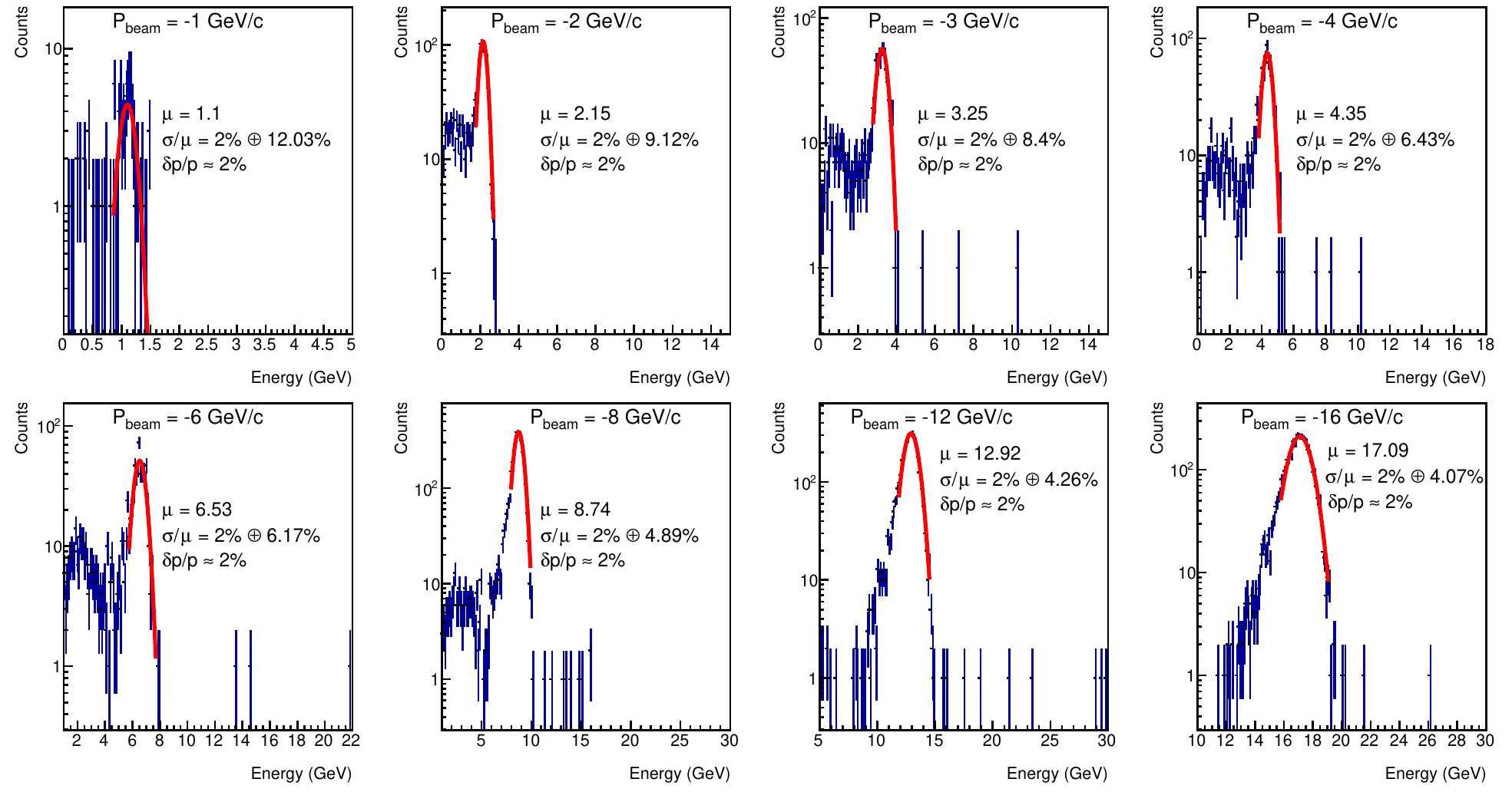}
  \caption{Cluster energy distribution of electron showers in the EMCal (blue points), for
  which the beam incident angle is 10 degrees and a 0.5~$\times$~0.5~cm$^2$ beam cross section is
  selected at the center of one EMCal tower. The central tower and most near-by tower are
  produced at UIUC. For each panel, data for one choice of beam energy is selected as
  shown in the title, and the energy resolution prior to unfolding a beam momentum spread
  ($\delta p/p \approx 2\% $) is extracted with a Gaussian fit at the electron peak (red
  curve). Low energy tails stemming from multi-particle background are excluded from the fit.}
  \label{fig:5by5_fits_scan2}
\end{figure*}

\begin{figure*}[htb!]
  \centering
  \includegraphics[width=.8\textwidth]{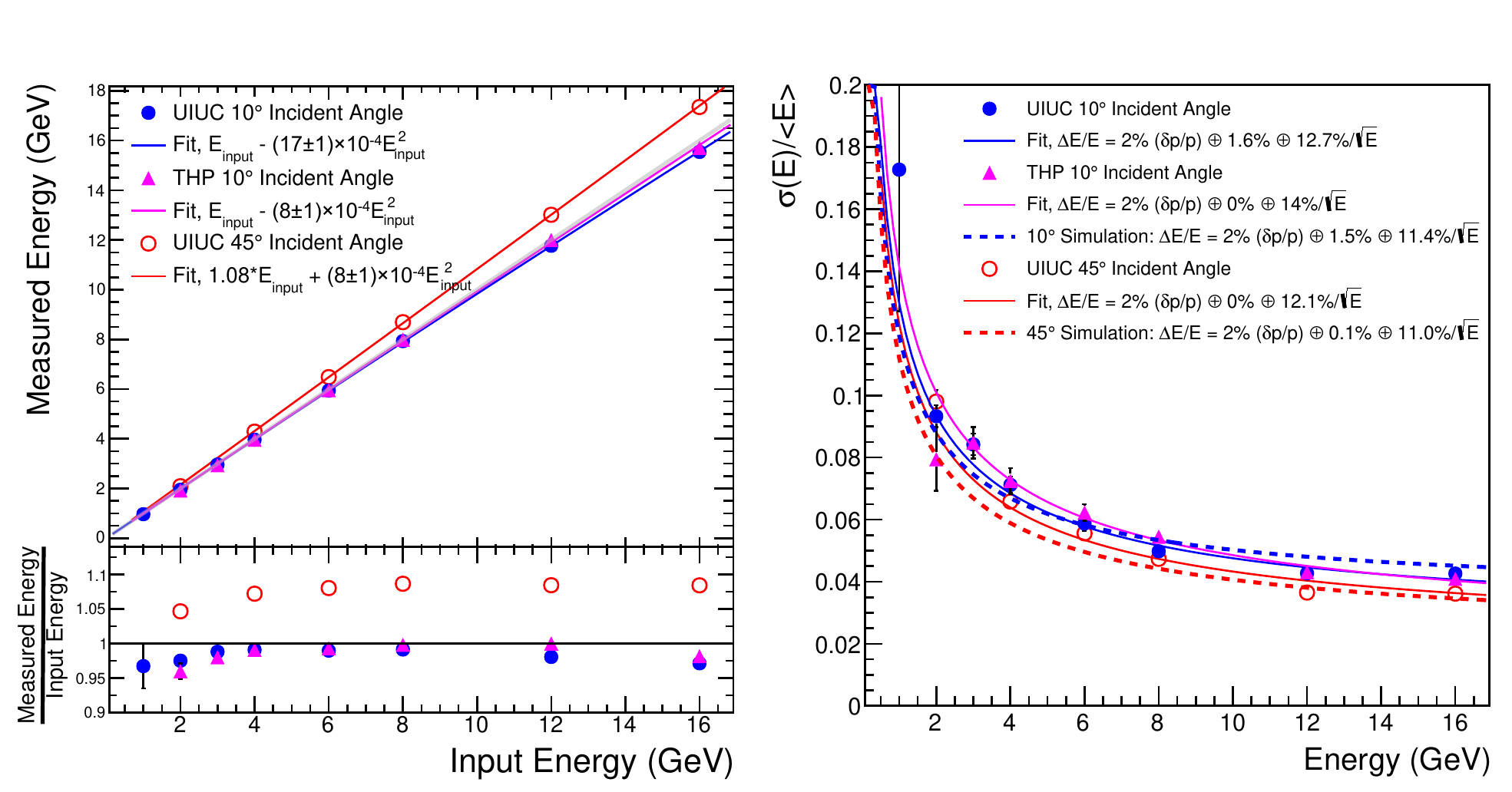}
  \caption{Linearity and resolution of electron showers in EMCal towers produced at UIUC and THP,
  for which a 1.0~$\times$~0.5~cm$^2$ beam cross section is selected at the center of one EMCal
  tower. The beam incident angles are 10 degrees (blue) and 45 degrees (red). Data
  (points) are fit with linear (left solid curves) and
  $\Delta E/E = \sqrt[]{a^2 + b^2/E}$ function with results labeled on plot (right solid
  curves), which are compared with simulation (dashed curves). A beam momentum spread
  ($\delta p/p \approx 2\% $) is unfolded and included in the resolution.  
  }
  \label{fig:5by5_fits_scan2_energyres}
\end{figure*}

\begin{figure*}[htb!]
  \centering
  \includegraphics[width=.8\textwidth]{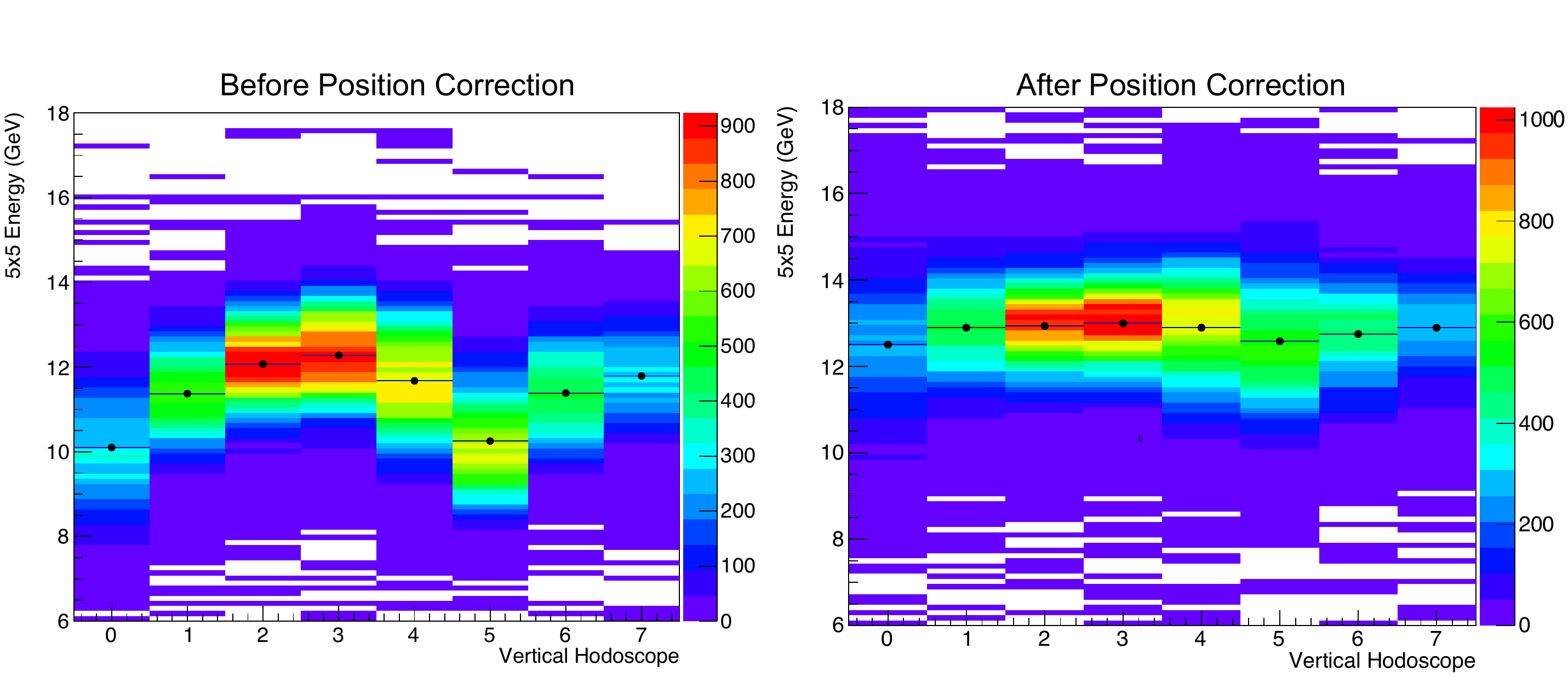}
  \caption{Cluster energy vs. vertical hodoscope in the EMCal towers produced at UIUC
  before and after the position-dependent energy correction is applied.
  The beam energy shown is 12~GeV with an incident angle of 10 degrees. Data is shown
  prior to unfolding a beam momentum spread
  ($\delta p/p \approx 2\% $). }
  \label{fig:5by5_HODOperformance}
\end{figure*}

\begin{figure*}[htb!]
  \centering
  \includegraphics[width=.75\textwidth]{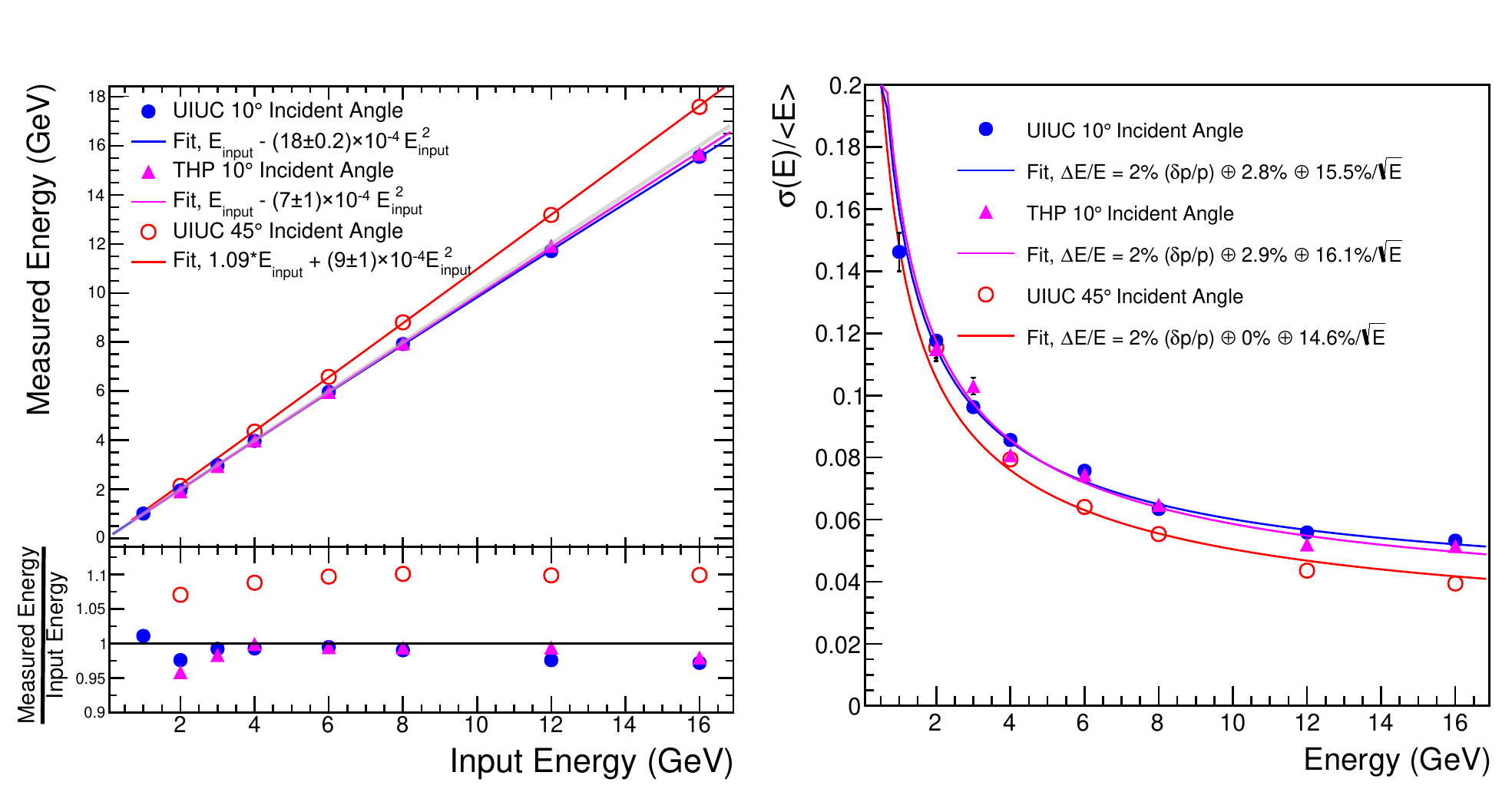}
  \caption{Linearity and resolution of electron showers in EMCal towers produced at UIUC and THP,
  for which a 2.5~$\times$~2.5~cm$^2$ beam cross section is selected and matches the area of one
  EMCal tower. The beam incident angles are 10 degrees (blue) and 45 degrees (red). Data
  (points) are fit with linear (left solid curves) and $\Delta E/E = \sqrt[]{a^2 + b^2/E}
  $ function with results labeled on plot (right solid curves). A beam momentum spread
  ($\delta p/p \approx 2\% $) is unfolded and included in the resolution.
  }
  \label{fig:5by5_fits_Tower21AverageHODO}
\end{figure*}

The electromagnetic energy resolution for the EMCal is obtained using the electron component of the test
beam, which is selected using the Cherenkov detectors, tuned to produce a signal for electron
events, but not for hadrons and muons.  However, due to multiple particle events as
discussed in Section~\ref{sec:testbeam}, the lower energy electron events still contain a
fraction of hadrons and muons, which are further rejected using the EMCal energy response.
For each event, the calibrated EMCal tower energy is summed within a 5~$\times$~5 tower
cluster centered around the tower with the maximum energy. When selecting a 0.5~$\times$~0.5~cm$^2$ beam cross section around the center of one tower, the 5~$\times$~5 tower
cluster energy is histogramed in
Figure~\ref{fig:5by5_fits_scan2}. The center of the tower is determined by selecting the hodoscope position with the highest average energy response in the EMCal.
The mean energy and spread of the EMCal response at each
beam energy is extracted with a Gaussian function fit at the electron peak 
. A beam momentum spread ($\delta p/p \approx 2\% $) is quadratically subtracted 
from $\sigma/\mu$ of the fit, in order to unfolded beam momentum spread from the relative energy resolution. The Gauss function parameter of $\mu$ and energy resolution from each fit are plotted against the nominal beam energy as linearity and resolution.
Two types of electron responses are studied:

\begin{itemize}

\item The energy resolution for showers located at the center of one tower,
which is a test of the intrinsic performance of the W/SciFi sampling structure
with minimal sensitivity to the light collection uniformity and tower edge effects.  With
a 1.0~$\times$~0.5~cm$^2$ beam hodoscope selection around the center of one tower, the
linearity and resolution are shown in Figure~\ref{fig:5by5_fits_scan2_energyres}
for SPACAL towers produced at UIUC and THP, respectively. At
a 10 degree incident angle, the performance of the UIUC and THP SPACAL towers 
appear qualitatively
comparable
with each other and with that of simulation, producing a resolution of
$\Delta E/E~=~1.6\%\oplus~12.7\%/\sqrt{E}$ after unfolding the beam momentum spread.
At a 45 degree beam incident angle, the resolution is found to be
$\Delta E/E~=~12.1\%/\sqrt{E}$ (with a small constant term when compared with the fit uncertainty) after unfolding the beam momentum spread.

\item Resolution with a beam cross section selection of 2.5~$\times$~2.5~cm$^2$,
which matches the full cross section of one SPACAL tower and is more relevant for the EMCal performance in sPHENIX.  The energy response of the
EMCal depends on the hit position of the incoming particle, which mainly stems from the
non-uniformity of light collection on the tower light guide and the absorber skin of SPACAL blocks as discussed in Section~\ref{sec:emcal}. 
The absorber skin of SPACAL blocks also leads to lower average sampling fraction when compared with that for the center of the block, and therefore, worse statistical term in the energy resolution. 
A position dependent energy scale correction is
applied to the current data based on the two-dimensional beam position as measured using a
0.5~$\times$~0.5~cm$^2$ hodoscope selection. Figure~\ref{fig:5by5_HODOperformance} shows the performance of
position-dependent energy correction, which clearly reduces the variation of the EMCal energy response. 
Figure~\ref{fig:5by5_fits_Tower21AverageHODO}
shows the result linearity and resolution for a sum of all electron events 
within a 2.5~$\times$~2.5~cm$^2$ beam cross section
after this correction is applied to the EMCal blocks produced at UIUC
and THP, respectively.  The EMCal resolution after unfolding the beam momentum spread is
$\Delta E/E~=~2.8\%\oplus~15.5\%/\sqrt{E}$ at a 10 degree beam incident angle and
$\Delta E/E~=~14.6\%/\sqrt{E}$ (with a small constant term when compared with the fit uncertainty) at a 45 degree beam incident angle. 

\end{itemize}

For both Figure~\ref{fig:5by5_fits_scan2_energyres}~and~\ref{fig:5by5_fits_Tower21AverageHODO}, the linearity response at an incident angle of 45~degrees is approximately 10\% higher than at 10~degrees. This difference is expected, since at larger angles, the total energy of the shower is contained more in the narrow end of the SPACAL towers where the fiber density, and hence the sampling fraction is higher. For 2--3~GeV beam energies, the linearity deviates slightly from the perfect linearity due to the uncertainty in the actual beam energy from the nominal beam energy setting. This variation was also observed with the Pb-glass calorimeter. At higher energies, the measured energy deviates systematically below the nominal beam energy due to back leakage from the calorimeter modules.
With the second order polynomial fit, the maximum deviation of the linearity parameterizations from data is within $5\%$ for beam energy larger than 4~GeV.

\begin{figure}[htb!]
\centering
\includegraphics[width=1.0\linewidth]{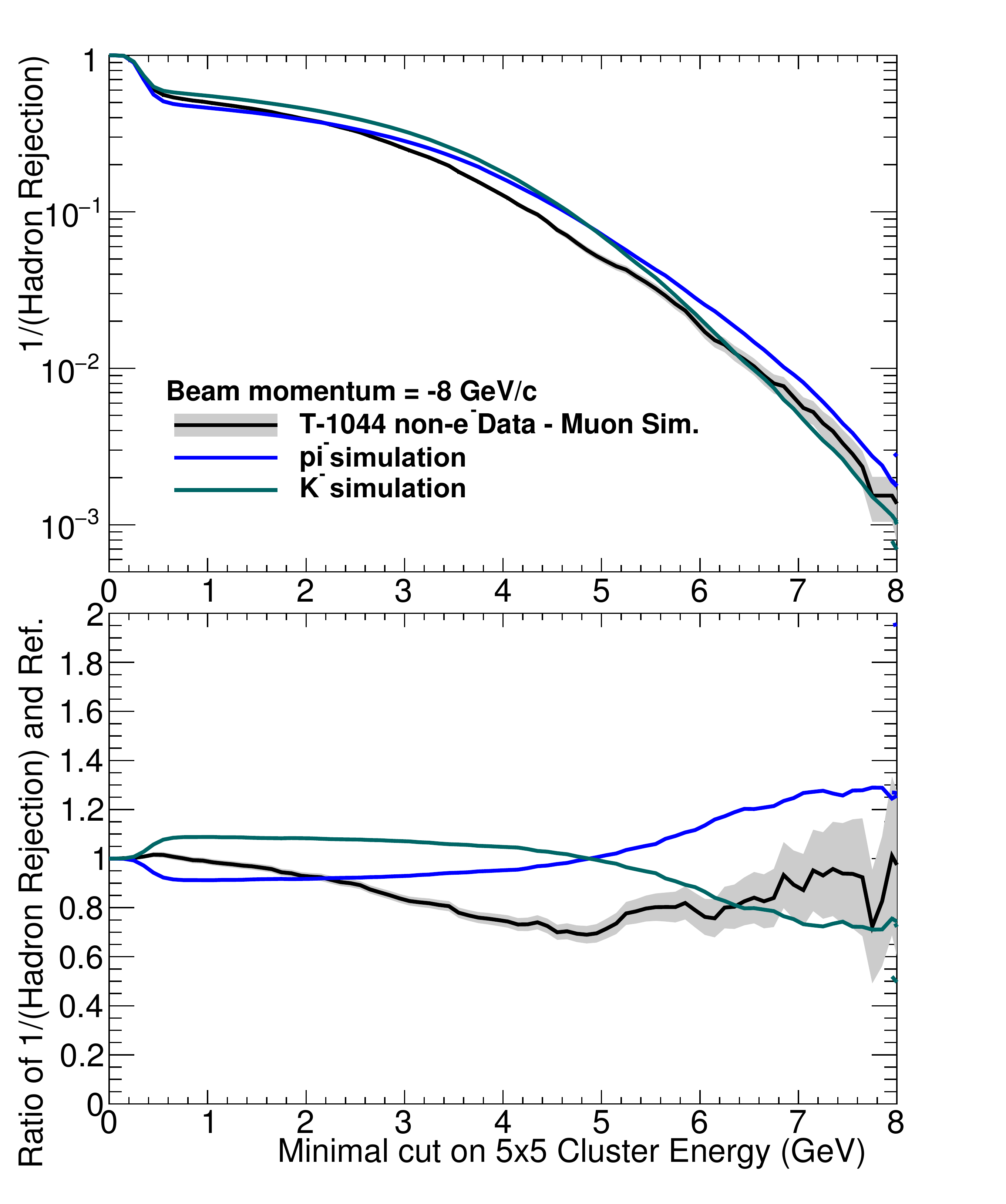}
\caption{\label{fig:Result_hadron_rejection}
Hadron rejection plotted against minimal cuts on 5~$\times$~5 tower cluster energy for negatively-charged
beam with momentum of 8~GeV/$c$. The T-1044 hadron data (black curve with
statistical uncertainties in gray), which are non-electron data with the expected
muon contribution subtracted, are compared with $\pi^-$ and $K^-$ simulated curves
. The beam momentum spread of around 2\% is present in both data
and simulation.}
\end{figure}

An important function of the EMCal in sPHENIX is to provide electron identification and
hadron rejection for charged tracks.
The hadron rejection factor is quantified as the ratio between the total number of incident hadron
and the subset of those with an EMCal cluster passing a minimal $E/p$ cut, as measured and simulated in
Figure~\ref{fig:Result_hadron_rejection}.
The accepted beam impact points cover an area of 2.5~$\times$~2.5~cm$^2$
as tagged by the hodoscope detector. 
The hadronic beam particles 
are selected by requiring no activity in the beam-line Cherenkov detectors, which are
tuned to produce Cherenkov signals on electrons but not on hadrons and muons.  Based on
Figure~\ref{fig:beamprofile}, the expected muon component in the beam is simulated and
statistically subtracted from the cluster energy spectrum. 
We note that the muon simulation on EMCal is not directly validated with this test beam.
Therefore, this subtraction could lead to a component in the systematic uncertainty 
for the hadron rejection results, 
which is still to be investigated.
The resulting EMCal cluster
energy spectrum for hadrons is integrated from various cut values to the maximum energy  in order to estimate the number of hadron events with cluster energy larger than the cut. Its ratio to the total number of hadron events is plotted as inverse of the hadron rejection factor versus
minimal cluster energy cut as shown in Figure~\ref{fig:Result_hadron_rejection}. This hadron sample contains mainly $\pi^-$. The kaon
content is expected to be very small, about 1\% of beam content at higher momenta
(20--30~GeV$/c$)~\cite{Blatnik:2015bka}, and lower at lower momenta (4--12~GeV$/c$) due
to the decay of kaons in flight.
Nevertheless, for the completeness of the study, both $\pi^-$ and $K^-$ are simulated
.
The result with beam momentum of 8~GeV$/c$ is shown in Figure~\ref{fig:Result_hadron_rejection} as a typical result,
while this study is performed in a negatively charged hadron beam momenta of 4, 8 and 12~GeV$/c$.
\subsection{HCal Calibration}

\begin{figure*}[!hbt]
  \begin{center}
    
    \subfloat[]{\includegraphics[width=0.45\textwidth]{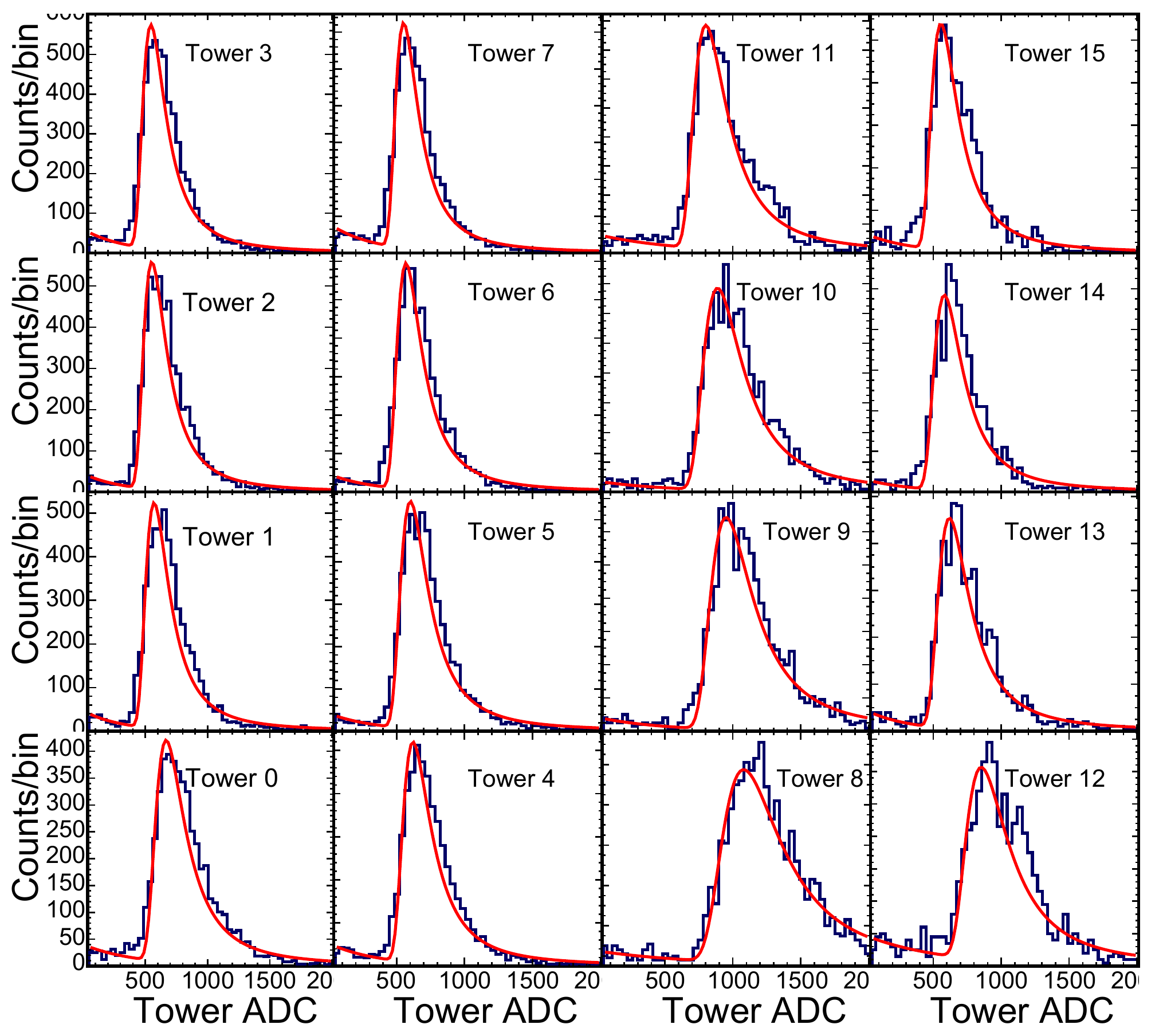}}
    \subfloat[]{\includegraphics[width=0.375\textwidth]{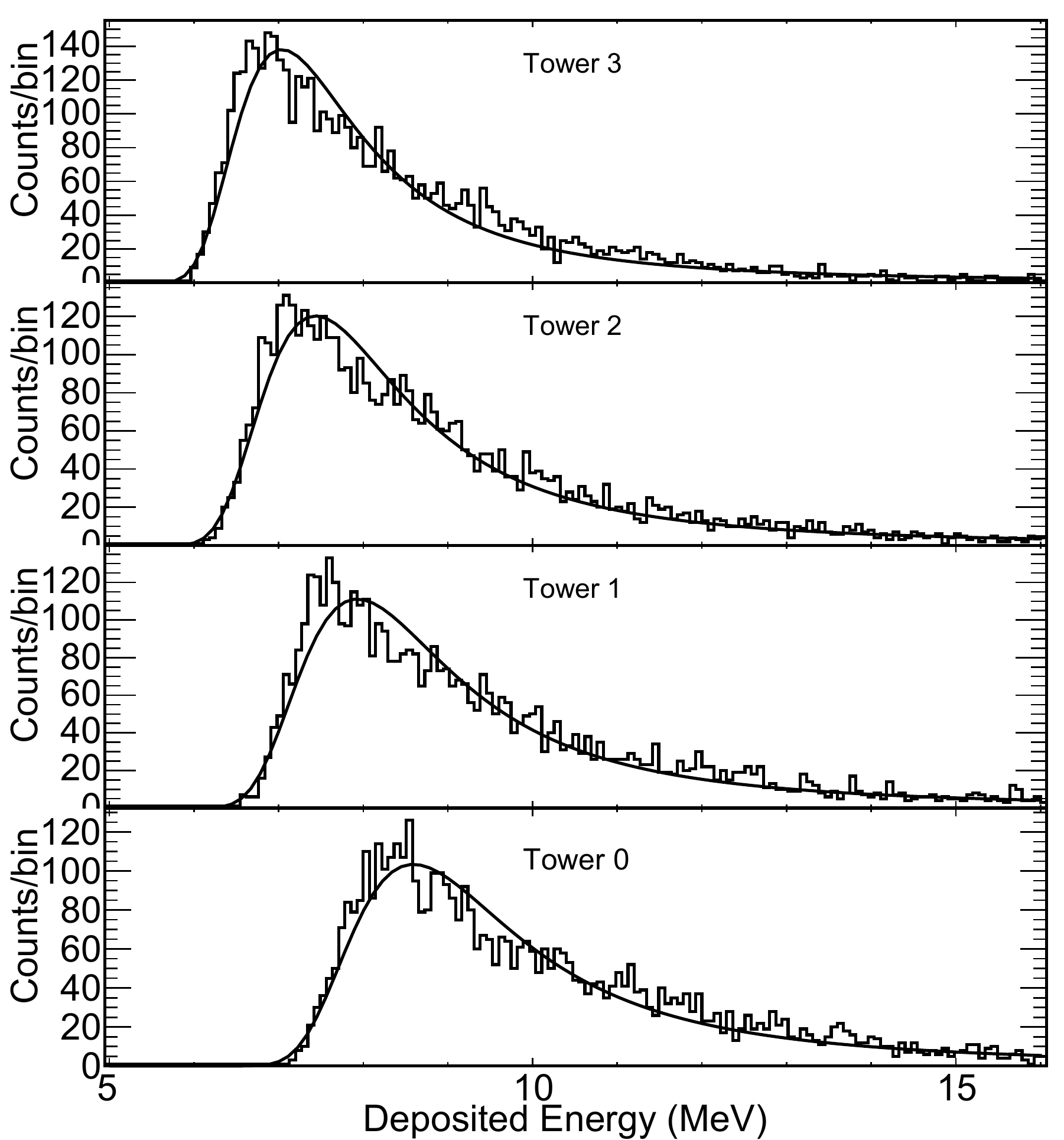}}
    
  \end{center}
  \caption{Tower to tower calibration for inner and outer HCal was done with cosmic muons.
  (a) Measured raw ADC spectra of cosmic ray muon events in the inner HCal. (b) Inner HCal
  cosmic muon energy deposition in simulation in one column. Muons were simulated at 4 GeV
  moving from the top to bottom. Energy depositions in the bottom towers are higher due to
  the tilted plate design where muons have to go through a longer path through the
  scintillating tiles.
  }
  \label{fig:hcalin_calib}
\end{figure*}

The initial HCal calibration was performed using cosmic MIP events in order to equalize the response of each tower.
A set of cosmic MIP events was
recorded prior to the test beam data taking in order to calibrate the detector. The cosmic
MIP events were triggered with scintillator paddles positioned at the top and bottom of
the HCal (in the $\phi$ direction as seen from the interaction point). In each run, four vertical towers are scanned from top to bottom (e.g. Tower 0--3 in
Figure~\ref{fig:hcalin_calib}). This yields eight individual runs in order to fully calibrate
both the inner and outer HCal sections. Figure~\ref{fig:hcalin_calib}~(a) shows the ADC distributions
in the 4~$\times$~4 inner HCal towers. Each spectrum is fit with the sum of an exponential and a Landau distribution, where the exponential function corresponds to the background and the Landau
function represents the MIP events. As seen in the figure, the background component is
relatively small. Clear cosmic MIP peaks are observed in all towers.

The corresponding simulation of cosmic muons is performed with 4~GeV muons 
(corresponding to the mean muon energy
at sea level) moving from the top to bottom of the HCal prototype with the 
\geant
setup discussed in Section~\ref{sec:simulations}. Figure~\ref{fig:hcalin_calib}~(b) shows energy deposition in only one column of towers. The mean energy deposited by the cosmic muons in each tower is approximately 8~MeV for
the inner HCal. Because of the tilted plate design, towers at the bottom of the inner HCal
have more deposited energy than the top ones. This feature was first observed in data and
then confirmed by the simulations. 
This simulation was used to calibrate the ADC signal height in
each tower, $I(ch)$, towards the corresponding energy deposition in the test beam: 

\begin{equation}\label{eq:hcal_cosmic_calib}
E(ch) = I(ch) \frac{E^{cosmic}_{dep}(ch)}{E^{ADC}_{dep}(ch) \times SF(muon)},
\end{equation}
where $E^{cosmic}_{dep}(ch)$ is the total deposited energy extracted from the \geant
cosmic simulations, $E^{ADC}_{dep}(ch)$ is the ADC signal height measured from cosmic data, and
$SF(muon)$ is the muon sampling fraction.
Note that the final energy scale is not set by the cosmic calibration alone but rather by a balancing procedure described in the following sections. Additional studies to further validate the cosmic simulations are still underway.

\subsection{HCal Standalone Measurements}

\begin{figure}[!hbt]
  \centering
  \includegraphics[width=1.\linewidth]{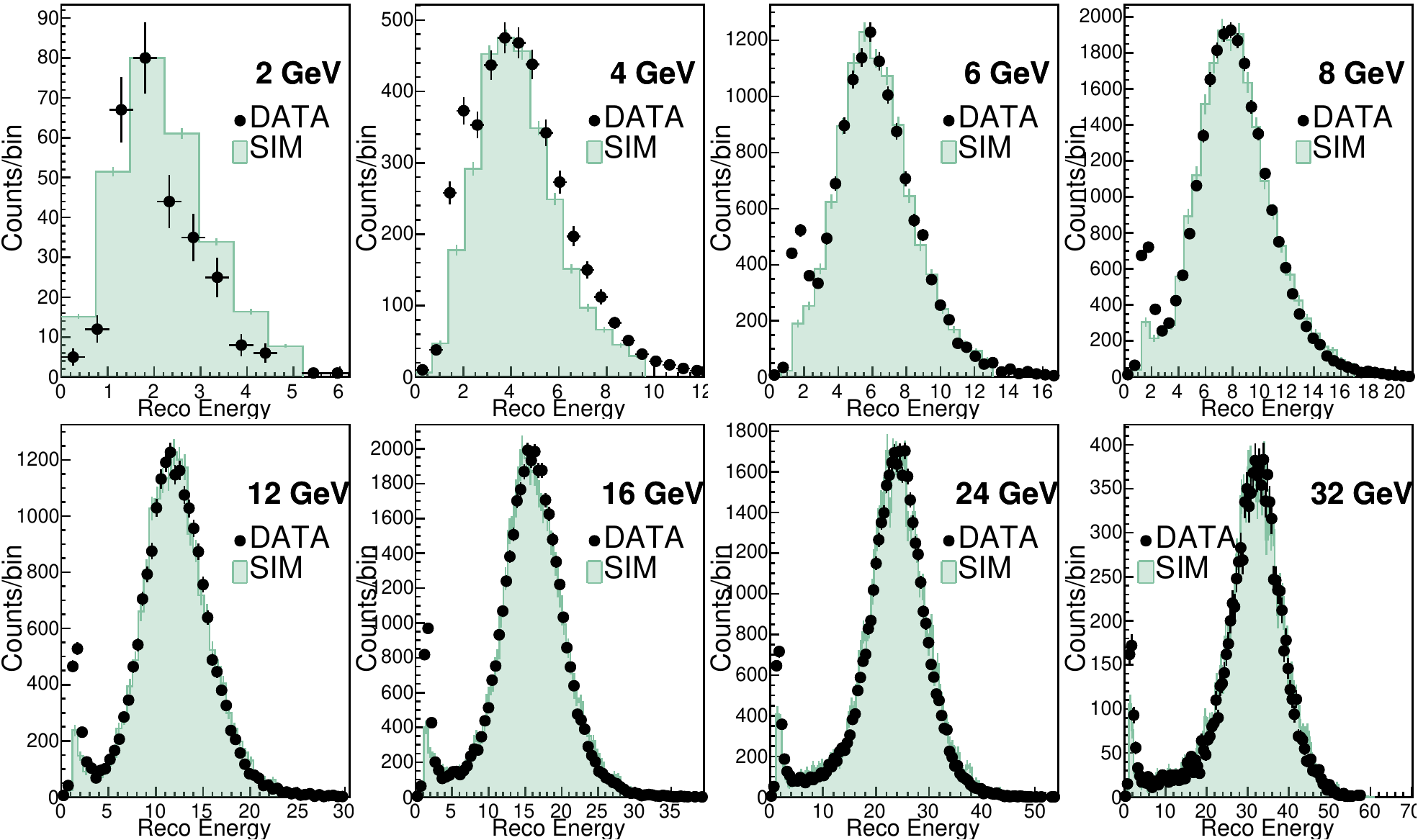}
  \caption{Hadron reconstruction in standalone HCal setup. Calibrated $4\times4$ tower
  energies were added together from the inner and the outer HCal. The simulation is shown by the
  filled histogram and the solid points are the data. 
  The peak
  at the lower energies in the data corresponds to the small fraction of muon events that
  pass through the HCal leaving only the minimum ionizing energy, which were not simulated.}
\label{fig:HCAL_standalone_signal}
\end{figure}

\begin{figure*}[!hbt]
\begin{center}
\subfloat[]{\includegraphics[width=0.40\textwidth]{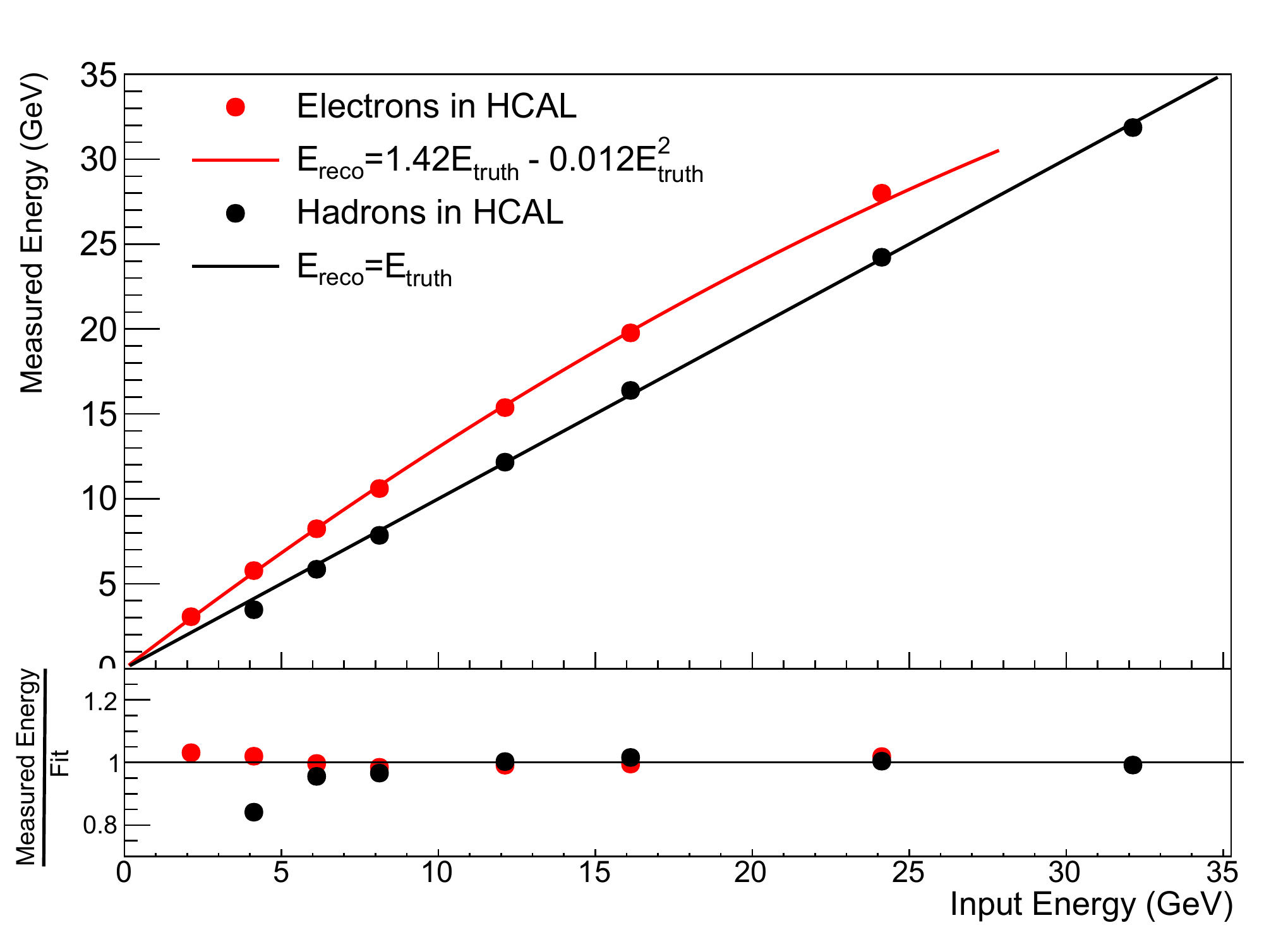}}
\qquad
\subfloat[]{\includegraphics[width=0.45\textwidth]{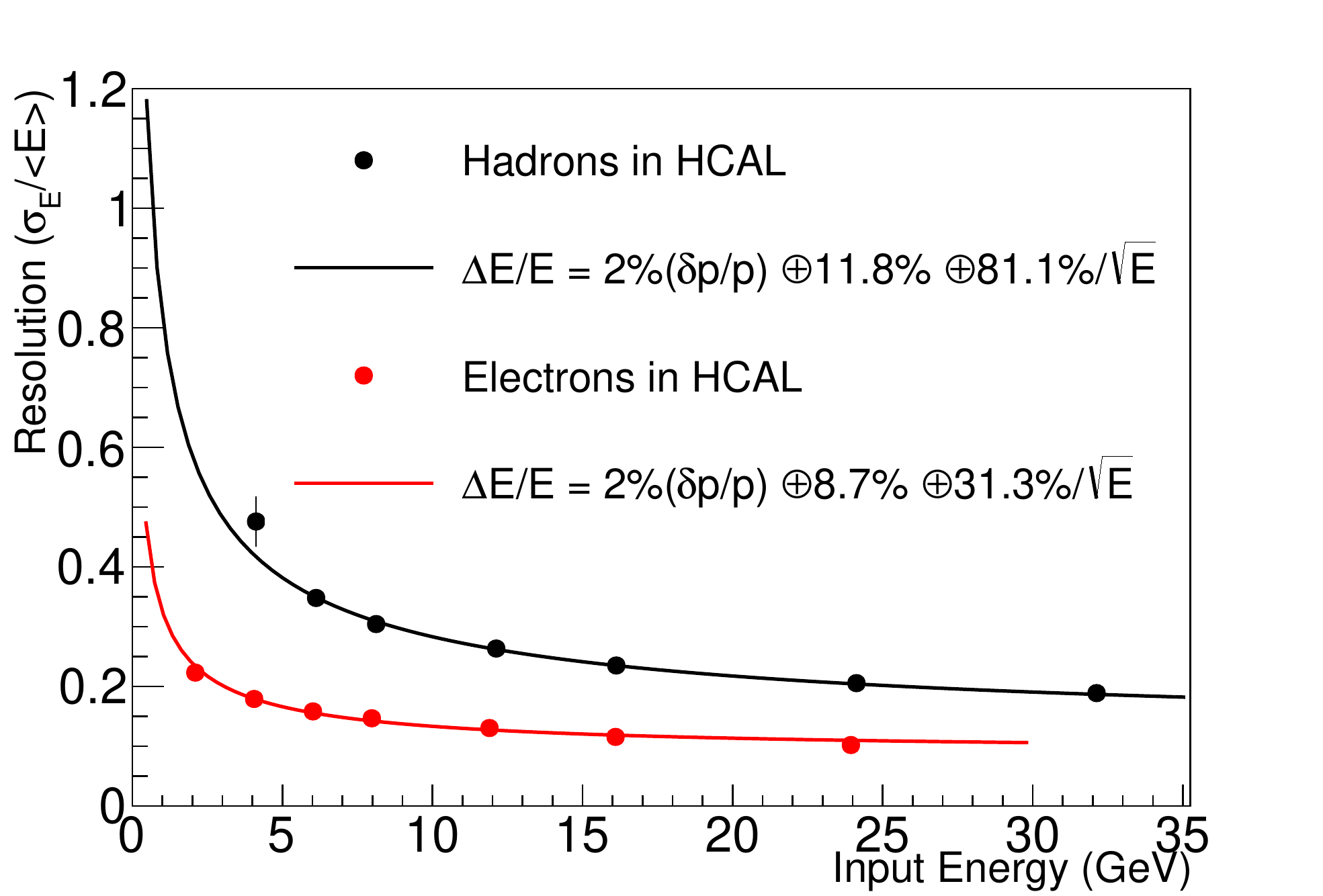}}
\end{center}
\caption{HCal standalone measurements without the EMCal in-front.
(a) HCal linearity for electrons and hadrons.
The lower panel shows the ratio of reconstructed energy and the fits. 
(b) Corresponding HCal resolution for hadrons and electrons.
The beam momentum spread ($\delta p/p \approx 2\%$) is unfolded and included in the resolution calculation.}
\label{fig:HCAL_Standalone}
\end{figure*}

\begin{figure}[!hbt]
  \centering
  \includegraphics[width=\linewidth]{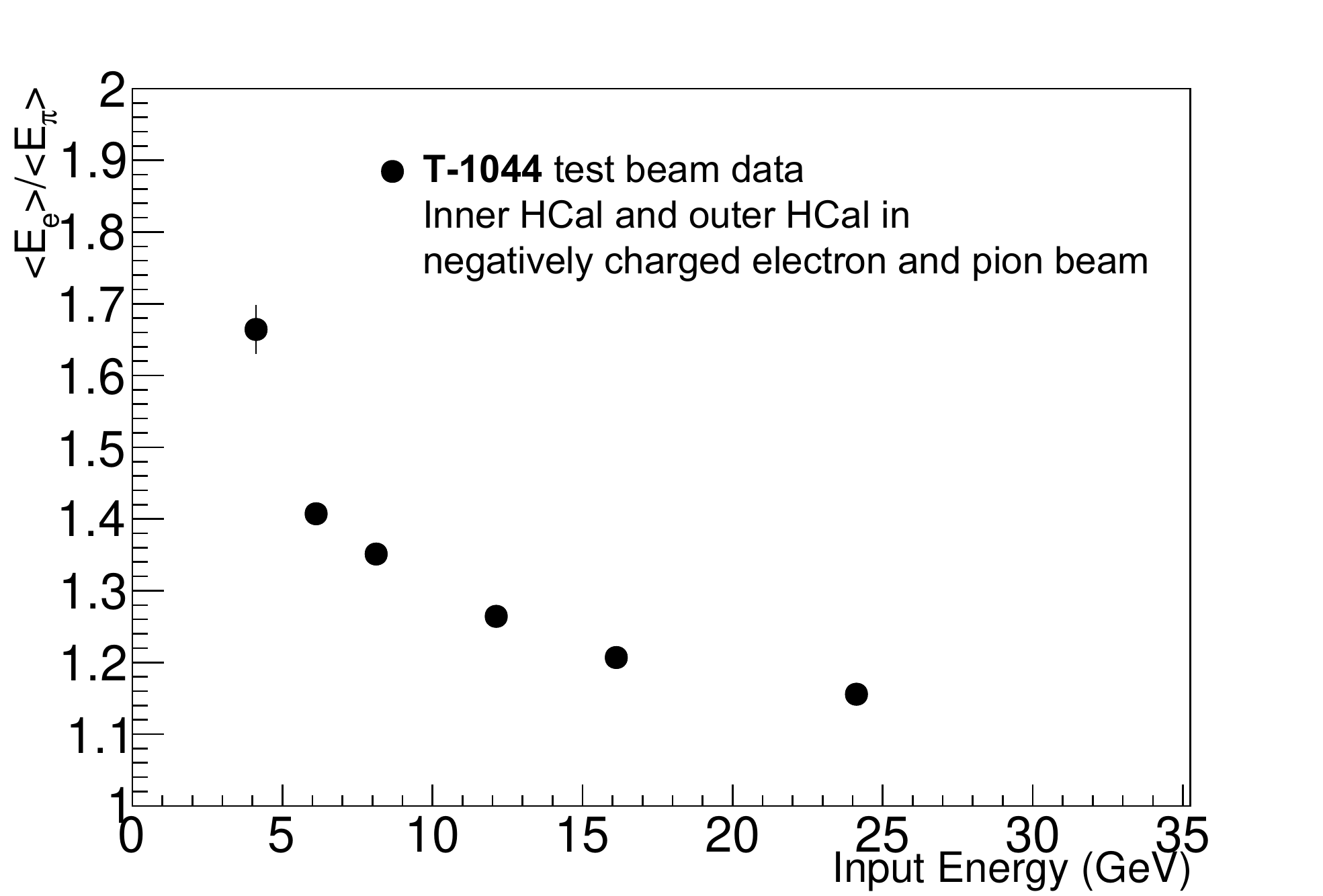}
  \caption{HCal $\langle E_{e}\rangle/\langle E_{\pi}\rangle$ response 
  as function of beam energy.
  }
  \label{fig:HCAL_eoverpi}
\end{figure}

The HCal standalone data are collected with only inner and outer sections of HCal in the
beam line and no EMCal in front.
In this configuration, electromagnetic showers generally start
earlier in the calorimeter and deposit most of their energy in the inner HCal. The
hadronic showers, however, are typically deeper than the electromagnetic
showers and deposit most of their energy in the outer HCal. The beam is adjusted to be in the middle of the
prototypes in order to maximize the hadron shower containment in the 4~$\times$~4 inner and outer
HCal towers.
Data were collected with a negatively charged particle beam with energies between 2 GeV and 32 GeV, which contain mainly electrons and pions as shown in Figure~\ref{fig:beamprofile}. Electron and pion events were tagged using
the two beamline Cherenkov counters. Hodoscope and veto cuts were applied depending on the
beam location, similar to the EMCal analysis, but found no large dependence of the energy
resolution on the beam position.
Both high and low gain signals from the HCal towers were collected but only
low gain channels are used for analysis.

The energy from all of the towers of both the inner and outer HCal are summed to determine the reconstructed energy:
\begin{equation}
E_{HCAL} = Gain_{inner} E_{inner}+ Gain_{outer} E_{outer}
\end{equation}
where $E_{inner}$, $E_{outer}$ are sum of the calibrated tower energy, $\Sigma_{ch} E(ch)$, within the inner and outer HCal, respectively.
The asymmetry between the two sections is defined as
\begin{equation}
A_{HCAL} = \frac{E_{inner}-E_{outer}}{E_{inner}+E_{outer}}.
\end{equation}
The gain calibration constants, $Gain_{inner}$ and $Gain_{outer}$, are determined in order to minimize the dependence of $E_{HCAL}$ on $A_{HCAL}$ and the deviation of $E_{HCAL}$ from the beam energy. The same gain calibration constants are used in analysis of all beam energies.

Figure~\ref{fig:HCAL_standalone_signal} shows the reconstructed hadron
energy in data and simulation. 
The beam momentum spread is not unfolded in both cases. 
At lower energies, hadron measurements are poor due to lower fractions of hadrons in the beam (Figure~\ref{fig:beamprofile}) as well as the increased beam
size. The peak at the lower energies in
the data corresponds to the small fraction of muons events that pass through the HCal
leaving only the minimum ionizing energy.
The corresponding hadron resolution and
linearity are shown in Figure~\ref{fig:HCAL_Standalone}. The data are fit with the function, $\Delta E/E = \sqrt{a^2+b^2/E}$, as labeled on the plot. A beam momentum spread ($\delta p/p \approx 2\%$) is unfolded and included in the resolution calculation. The hadron energy resolution
follows an empirical formula $11.8\%~\oplus~81.1\%/\sqrt{E}$ 
with 
a p-value of 0.37.
The HCal was calibrated for hadronic showers and then used to measure electron showers. The electron resolution for the standalone HCal is
$8.1\%~\oplus~31.3\%/\sqrt{E}$. This demonstrates the HCal's ability to assist the EMCal by
measuring the electron energy leaking from the EMCal into HCal.

As  seen in Figure~\ref{fig:HCAL_Standalone} (a), the hadron energy response 
can be qualitatively
described by a linear fit where the reconstructed energy is the same as the input energy. The
bottom panel shows the ratio between the reconstructed energy and the fit. The 4~GeV
hadron measurement is poor because the hadron peak is difficult to distinguish
 from the muon MIP peak as seen in
Figure~\ref{fig:HCAL_standalone_signal} due to their proximity.  
The electrons can be described with a second order polynomial where the second order coefficient is $0.012\pm8.8\times10^{-5}$.
The ratio between the electron data and the fit are also shown in the bottom panel of Figure~\ref{fig:HCAL_standalone_signal}.
Furthermore, Figure~\ref{fig:HCAL_eoverpi} shows the HCal $\langle E_{e}\rangle/\langle E_{\pi}\rangle$
response 
as function of the beam energy, which is always higher than 1 and is energy dependent.

\subsection{Hadron Measurement with sPHENIX Configuration}

\begin{figure}[!hbt]
\centering
\includegraphics[width=\linewidth]{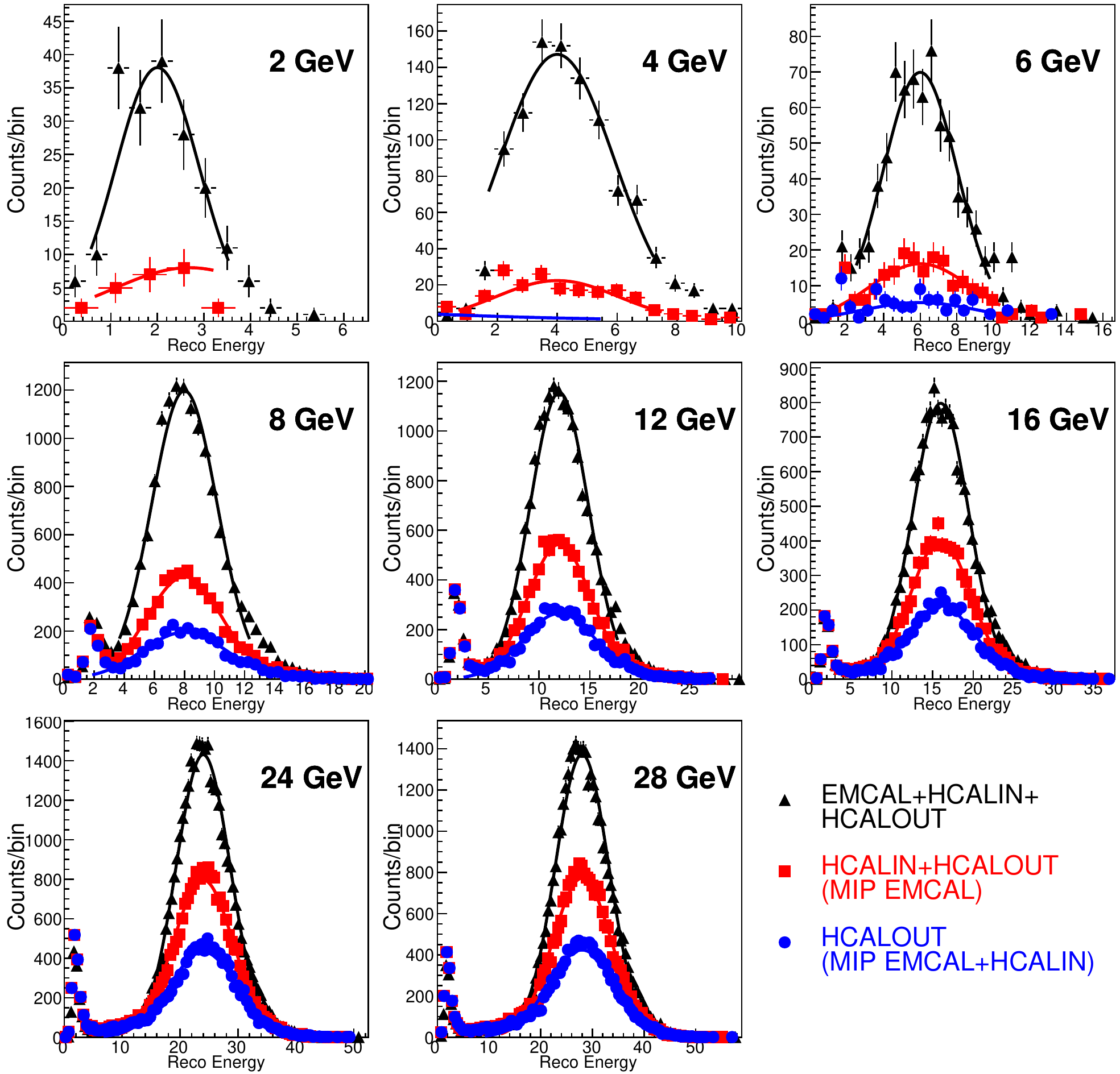}
\caption{Hadron energy measurement with combined EMCal+HCal detector.
Events were sorted into three categories:
1) HCALOUT where particles pass through the EMCal and inner HCal and then shower in the outer HCal;
2) HCALIN+HCALOUT where particles pass through the EMCal and then shower in either HCal;
3) EMCAL+HCALIN+HCALOUT which includes all showers irrespective of their starting position.}
\label{fig:hadron_signal_categories}
\end{figure}

\begin{figure*}[!hbt]
\begin{center}
\subfloat[]{\includegraphics[width=0.46\textwidth]{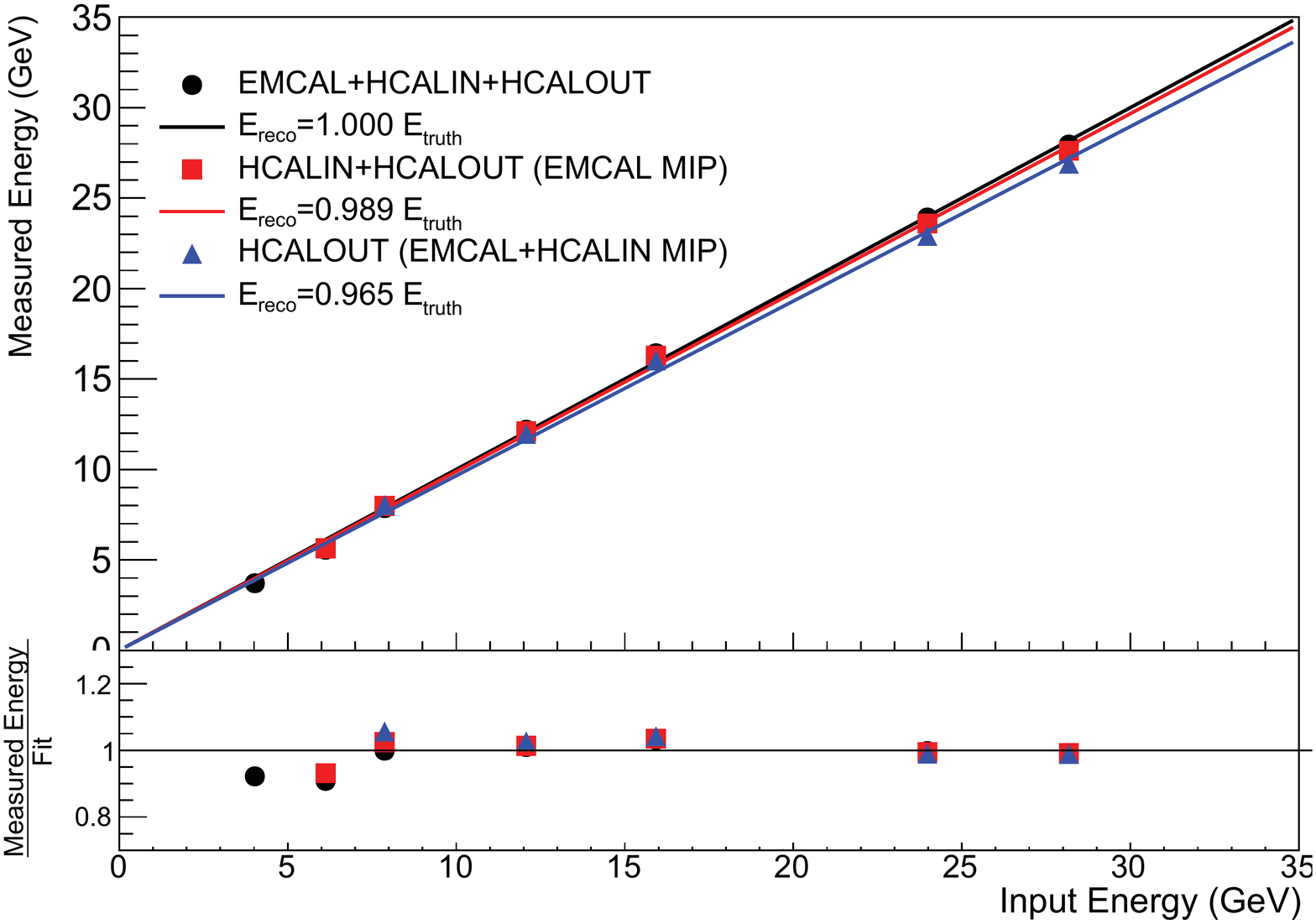}}
\quad
\subfloat[]{\includegraphics[width=0.49\textwidth]{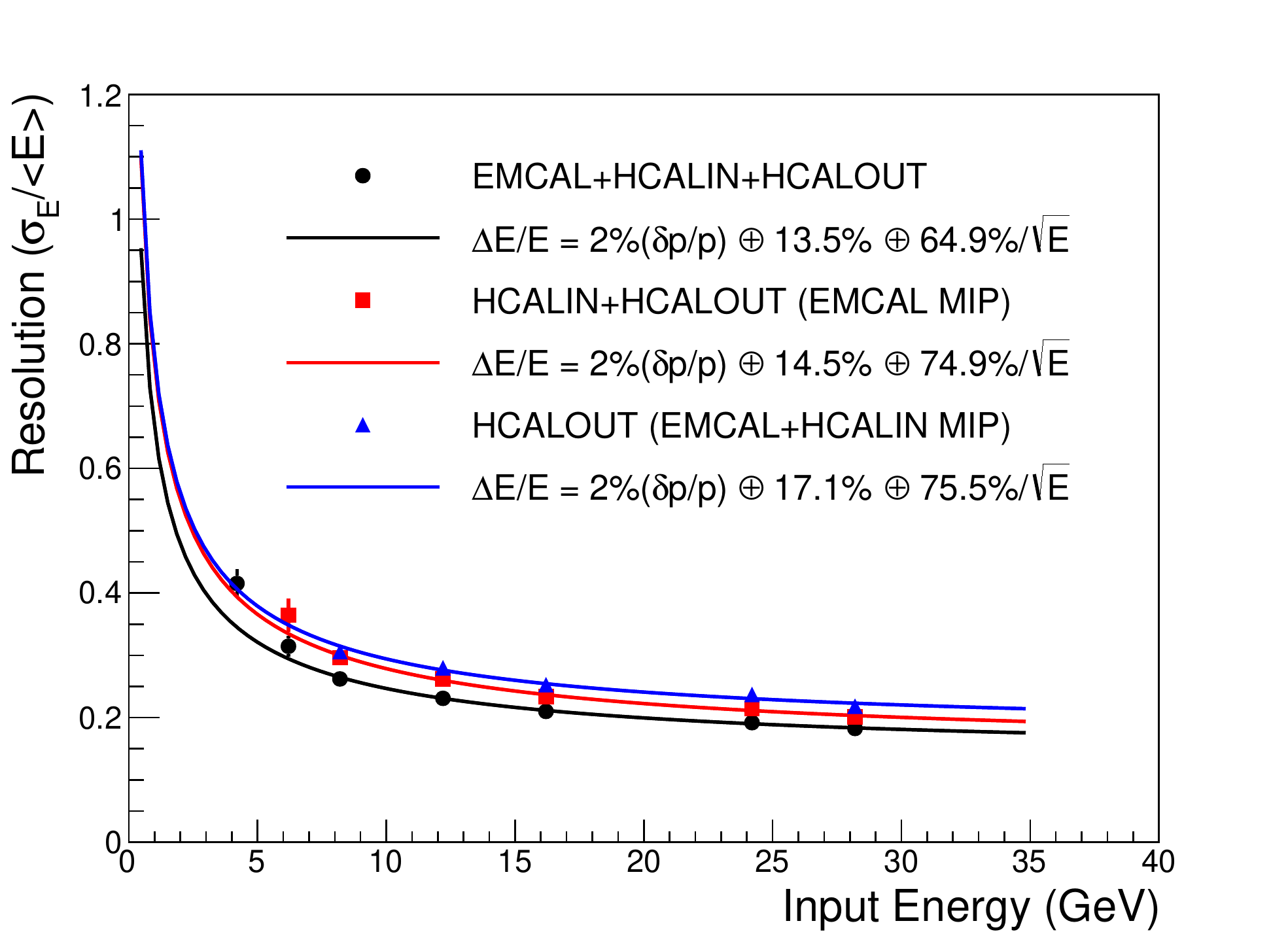}}
\end{center}
\caption{Hadron (a) linearity and (b) resolution measured with combined EMCal+HCal
(sPHENIX configuration) detector setup. Three sets of data points corresponds to the event
categories shown in Figure~\ref{fig:hadron_signal_categories}. The bottom panel of (a)
shows the ratio of the measured energy and corresponding fits.
}
\label{fig:hadron_resolution}
\end{figure*}

The full hadron measurement is done with the sPHENIX configuration, which includes all
three segments of calorimeters including the EMCal in front of the HCal. In this
configuration the total energy will be reconstructed by summing up the digitized data from
both the EMCal and the HCal.  The development of hadronic showers is a complicated process
with significant fluctuations of the reconstructed energy compared to electromagnetic
showers. Distinguishing the shower starting position helps to understand the
longitudinal shower development fluctuations.
Therefore, in this analysis, the events are sorted into three 
inclusive
categories depending on their longitudinal shower
profile:
\begin{itemize}
\item \textit{HCALOUT}: Events where hadrons pass through the EMCal and inner HCal and primarily shower in the outer HCal alone or pass through the full calorimeter
      system without showering. These events are shown as the blue points in
      Figure~\ref{fig:hadron_signal_categories}.

\item \textit{HCAL}: Events where hadrons pass through the EMCal. In these events, hadron showers
      start in the inner HCal, or the outer HCal, or pass through all three calorimeters.
      These events are shown as red points in Figure~\ref{fig:hadron_signal_categories}.

\item \textit{FULL}: This represents all hadrons irrespective of when they start showering.  They
      are shown as black points in Figure~\ref{fig:hadron_signal_categories}.  These
      include hadron showers that start in the EMCal, inner HCal, outer HCal, or
      pass through all three calorimeter systems.

\end{itemize}

These event categories help diagnose each section of the calorimeters independently as
well as understanding of the leakage variations, shower containment and longitudinal
fluctuations depending on their starting position. 
The EMCal energy was balanced with respect to
the HCal in a similar way by changing the gain factors prior to summing them into the total reconstructed energy, $E_{Total}$ according to:
\begin{equation} \label{eq:e_total}
E_{Total} = Gain_{EMCal} E_{EMCal} + Gain_{HCal} E_{HCal}
\end{equation}
where $E_{EMCal}$ is the sum of the calibrated tower energy within the EMCal.
The procedure of adjusting the gain coefficients is very similar to the one 
described in the last section for balancing inner vs outer HCal: $Gain_{EMCal}$ and $Gain_{HCal}$ are adjusted in order to minimize the dependence of $E_{Total}$ on $(E_{EMCal} - E_{HCal})/(E_{EMCal} + E_{HCal})$
and the deviation of $E_{Total}$ from the beam energy. 
The FULL shower sample is used for this calibration. 
The HCal gains were held fixed with respect to the beam energy, 
while the EMCal gain was adjusted for each beam energy separately.
Figure~\ref{fig:hadron_signal_categories} shows total reconstructed energy as obtained in Equation~(\ref{eq:e_total})
for each of the three event categories and various beam energies.
The peaks at the lower energy correspond to the small fractions of muon events that pass
through the calorimeters leaving only the minimum ionizing energy.

The corresponding hadron resolution is shown in Figure~\ref{fig:hadron_resolution} (b).
Data are fit in a similar manner with $\Delta E/E = \sqrt{(\delta p/p)^2 + a^2+b^2/E}$,
i.e. with a fixed beam momentum spread term of $\delta p/p \approx 2\%$
subtracted from the constant term in quadrature. HCALOUT showers that pass through the EMCal
and inner HCal have a resolution of $17.1\%~\oplus~75.5\%/\sqrt{E}$. 
The p-value for this fit is 0.0041.
HCAL showers that pass through
through the EMCal have a resolution of $14.5\%~\oplus~74.9\%/\sqrt{E}$ 
which gives a p-value of 0.016
. The combined resolution of
all the showers irrespective of their starting position (FULL) is
$13.5\%~\oplus~64.9\%/\sqrt{E}$ 
with a p-value of 0.0084
. The hadron resolution improves without the MIP cuts because it
reduces the overall shower fluctuations and leakages. 

The linearity is shown in Figure~\ref{fig:hadron_resolution} (a). The bottom panel shows
the ratio of the measured energy and the corresponding fits. 
As the FULL data sample is used to adjust the energy sum calibration as in Equation~(\ref{eq:e_total}), the linear fit coefficient of its measured energy to beam energy is $1$ by definition. The same gain factors were applied to the HCAL and HCALOUT shower categories. We 
qualitatively
observed their linearity slope slightly below 1, which could be due to slightly higher energy leakage in those event categories. 

\section{Conclusions}
\label{sec:conclusions}
A prototype of the sPHENIX calorimeter system was successfully constructed and tested at
the Fermilab Test Beam Facility with beam energies in the range of 1-32~GeV. The energy
resolution and linearity of the EMCal and HCal were measured as a combined calorimeter
system as well as independently.
The energy resolution of the HCal is found to be
$\Delta~E/E~=~11.8\%\oplus~81.1\%/\sqrt{E}$ for hadrons.
The energy resolution of EMCal for electrons is $1.6\% \oplus 12.7\%/\sqrt{E}$ for electromagnetic showers that
hit at the center of the tower and $2.8\% \oplus 15.5\%/\sqrt{E}$ without the position
restriction. Part of the EMCal position dependence of the shower response stems from the
non-uniformity of the light collection in the light guide, which will be a major focus of the next stage of detector research and development.
The combined hadron resolution of the full EMCal and HCal system for
hadrons is $13.5\% \oplus 64.9\%/\sqrt{E}$ and is consistent with the standalone HCal results.
All of these results satisfy the requirements of the sPHENIX physics program.
Simulation studies are progressing in parallel to support the research and development of these detectors.

\section*{Acknowledgments}


This document was prepared by members of the sPHENIX
Collaboration using the resources of the Fermi National Accelerator
Laboratory (Fermilab), a U.S. Department of Energy,
Office of Science, HEP User Facility. Fermilab is managed by
Fermi Research Alliance, LLC (FRA), acting under Contract
No. DE-AC02-07CH11359. The authors would like to thank
Dr. O. Tsai at UCLA for sharing his experience in developing
W/SciFi calorimeter modules and for providing the hodoscope
that was used in the test beam, also thank the technical staffs
of the University of Illinois at Urbana–Champaign (UIUC)
and the Brookhaven National Laboratory for assistance in
constructing the prototype detectors, and also thank the University
of Colorado Boulder for the technical assistance of
putting together the test stand and characterizing the tiles.
This work was carried out in part in the Frederick Seitz
Materials Research Laboratory Central Research Facilities,
UIUC.

\thanks{C.A.~Aidala and M.J.~Skoby are with the Department of Physics at the University of Michigan, Ann Arbor, MI 48109-1040.}

\thanks{V. Bailey, J. Blackburn, M.M. Higdon, S. Li, V.R. Loggins, M. Phipps, A.M. Sickles, P. Sobel,
  and E. Thorsland are with the Department of Physics at the University of Illinois Urbana-Champaign, Urbana, IL 61801-3003.}

\thanks{S. Beckman, R. Belmont, J.L. Nagle, and S. Vazquez-Carson are with the Department of Physics at 
the University of Colorado Boulder, Boulder, CO 80309-0390.}

\thanks{C. Biggs, S. Boose, M. Chiu, E. Desmond, A. Franz, J.S. Haggerty, J. Huang, E. Kistenev, J. LaBounty,
  M. Lenz, W. Lenz, E.J. Mannel, C. Pinkenburg, S. Polizzo, C. Pontieri, M.L. Purschke, R. Ruggiero, S. Stoll, A. Sukhanov,
  F. Toldo, and C.L. Woody are with Brookhaven National Laboratory, Upton, NY 11973-5000.}

\thanks{M. Connors is with the Department of Physics and Astronomy at the Georgia State University, Atlanta, GA 30302-5060 and the RIKEN BNL Research Center, Upton, NY 11973-5000.}

\thanks{X. He and M. Sarsour are with the Department of Physics and Astronomy at the Georgia State University, Atlanta, GA 30302-5060.}


\thanks{K. Kauder is with the Department of Physics and Astronomy at the Wayne State University, Detroit, MI 48201-3718 and Brookhaven National Laboratory, Upton, NY 11973-5000}

\thanks{J.G. Lajoie, T. Rinn, and A. Sen are with the Department of Physics and Astronomy at the Iowa State University, Ames, IA 50011-3160.}

\thanks{T. Majoros and B. Ujvari is with the Institute of Physics at the University of Debrecen, Debrecen, Hungary.}

\thanks{M.P. McCumber is with Los Alamos National Laboratory, Los Alamos, NM 87545-0001.}

\thanks{J. Putschke is with the Department of Physics and Astronomy at the Wayne State University, Detroit, MI 48201-3718.}

\thanks{J. Smiga is with the Department of Physics at the University of Maryland, College Park, MD 20742-4111.}

\thanks{P.W. Stankus is with Oak Ridge National Laboratory, Oak Ridge, TN 37830-8050.}

\thanks{R.S. Towell is with the Department of Engineering and Physics at the Abilene Christian University, Abilene, TX 79699-9000.}

\bibliographystyle{IEEEtran}
\bibliography{IEEEabrv,T1044bibfile}

\end{document}